\documentclass[12pt]{article}

\usepackage{hyperref}

\usepackage{epsf}
\usepackage{amsmath,amssymb}
\usepackage{latexsym}
\usepackage[pdftex]{graphicx} 
\usepackage{wrapfig}
\usepackage{latexsym}

\usepackage{color}
\usepackage{delarray,color}
\usepackage{fancybox}  
\newcommand {\beq}{\begin{align}}
\newcommand {\eeq}{\end{align}}

\newcommand{\be}{\begin{equation}}
\newcommand{\ba}{\begin{align}}
\newcommand{\ea}{\end{align}}
\newcommand{\ee}{\end{equation}}

\newcommand{\beqa}{\begin{align}}
\newcommand{\eeqa}{\end{align}}
\newcommand{\CR}{\nonumber \\}

\newcommand{\unit}{\hbox to 3.8pt{\hskip1.3pt \vrule height 7.4pt
    width .4pt \hskip.7pt \vrule height 7.85pt width .4pt \kern-2.4pt
    \hrulefill \kern-3pt \raise 3.7pt\hbox{\char'40}}}

\def\matt[#1,#2,#3,#4]{\left(%
\begin{array}{cc} #1 & #2 \\ #3 & #4 \end{array} \right)}

\newcommand{\ket}[1]{{\left| #1 \right\rangle}}
\newcommand{\bra}[1]{{\left\langle#1\right|}}
\usepackage{tabularx}

\catcode`\@=11
\@addtoreset{equation}{section}

\catcode`@=12
\relax

\begin{document}
%%%%%%%%%%%%%%%%%%%%%%%%%%%%%%%%%%%%%%%%%%%%%%%%%%%%%%%%%%%%%%%%%%%%%%%%
%\baselineskip 0.7cm

\begin{titlepage}

%% Set the number of the title with 0
\setcounter{page}{0}

%% change the footnote symbol
\renewcommand{\thefootnote}{\fnsymbol{footnote}}

\begin{flushright}
%{\tt 
YITP-21-37
%\\}
\end{flushright}

\vskip 1.35cm

\begin{center}
{\Large \bf 
Simple Bulk Reconstruction \\
in AdS/CFT Correspondence
}

\vskip 1.2cm 

{\normalsize
Seiji Terashima\footnote{terasima(at)yukawa.kyoto-u.ac.jp}
}

\vskip 0.8cm

{ \it
Yukawa Institute for Theoretical Physics, Kyoto University, Kyoto 606-8502, Japan
}

\end{center}

\vspace{12mm}

\centerline{{\bf Abstract}}

In this paper, we show that the bulk reconstruction in the
AdS/CFT correspondence is rather simple and has an intuitive picture,
by showing that the HKLL bulk reconstruction formula can be simplified.
%Indeed, a bulk local operator at a point $p$ is reconstructed from 
%the CFT primary fields integrated over a submanifold $M_p$ in the spacetime for CFT, where $M_p$ is
%the intersection of the light-cone of the point $p$ and the boundary of the AdS spacetime.
%Note that this is for the free bulk theory approximation around the AdS spacetime.
We also reconstruct the wave packets in the bulk theory from 
the CFT primary operators.
%The bulk local operator can be represented by a linear combination of these wave packets
%at a same spacetime point, but moving in different directions.
With these wave packets, we discuss the causality and duality constraints
and find our picture is only the consistent one.
Our picture of the bulk reconstruction can be applied to 
the asymptotic AdS spacetime.%, assuming the BDHM relation.
%In the CFT, this background is a non-trivial state which can has the bulk semi-classical description.
%In particular, we discuss that for the (single sided) black hole the bulk reconstruction is only possible
%for the spacetime outside the (streched) horizon.

\end{titlepage}
\newpage

\tableofcontents
\vskip 1.2cm 

\section{Introduction and summary}

The holographic principle \cite{holo, Susskind} is one of the most important concepts
in quantum gravity.
The asymptotic AdS spacetime is the ideal settings for realizing the holographic principle
and the AdS/CFT correspondence \cite{Maldacena} explicitly realizes it.
If the holographic principle is true, the bulk gravity theory is equivalent to 
the lower dimensional theory without gravity.

To understand this, the most important question will be 
how the states/operators in the bulk gravity theory are reconstructed from the states/operators in 
the lower dimensional theory without gravity.
Indeed, if we can understand this bulk reconstruction, we can also explain
how the bulk gravity theory emerges from the lower dimensional theory.

In the AdS/CFT correspondence, 
this bulk reconstruction was given %in \cite{HKLL} 
for the free bulk theory approximation around the AdS spacetime,
which is a large $N$ limit of the corresponding
conformal field theory (CFT) \cite{BDHM}-\cite{op}. 
Here  this CFT is realized as a gauge theory 
with a rank $N$ gauge group around the vacuum.
In particular, 
assuming the BDHM relation \cite{BDHM} which is 
an AdS/CFT dictionary as 
like the GKPW relation \cite{GKP, W},
the explicit formula of the reconstruction of the bulk local operator 
by the CFT primary field was given in \cite{HKLL} and is called HKLL bulk reconstruction.
This formula is given as an integration of CFT primary field 
over a region in the spacetime of CFT
with a weight called the smearing function.
Unfortunately, this formula is not simple and rather counter intuitive 
with the causality as we will explain later.

In this paper, we show that the bulk reconstruction in the
AdS/CFT correspondence is rather simple and has an intuitive picture,
by showing that the HKLL bulk reconstruction formula can be simplified.
Indeed,
a bulk local operator at a point $p$ is reconstructed from 
the CFT primary fields integrated over a submanifold $M_p$ in the spacetime for CFT, where $M_p$ is
the intersection of the light-cone of the point $p$ and the boundary of the AdS spacetime.
Note that this is for the free bulk theory approximation around the AdS spacetime.
%however, this picture is still hold for the perturbation theory around it.
Note also that this can be regarded as a direct consequence of 
the %backward-
time evolution using
the equations of motion of the free (light-like propagating) theory 
and the BDHM relation which identifies the bulk local operators on the boundary 
as the CFT primary fields.
Although this picture was already obtained in \cite{local} using the identification 
of the bulk and boundary operators in the energy eigen state basis \cite{op},
we mainly use the HKLL reconstruction formula in this paper 
partly because this formula is well-known.

To be more precise, the above picture of the bulk reconstruction is exact 
for $\Delta=d-1$ (or when  $\Delta$ is an integer), where $\Delta$ is the conformal dimension of the CFT primary field.
We expect that this picture is true for the energy momentum tensor and the conserved currents.
For other $\Delta$, the above picture is still valid for bulk local states/operators.
If we would like to reconstruct bulk non-local operators which is given by an integration 
of bulk local operators over a bulk region, 
we need the CFT primary fields at space-like separated boundary points, not just 
the ones at light-like separated points, for a generic $\Delta$.
In this paper, we concentrate on the bulk local states/operators.

We also reconstruct the wave packets in the bulk theory from 
the CFT primary operators.
The bulk local operator can be represented by a linear combination of these wave packets
at the same spacetime point, but moving in different directions.
With these wave packets, we discuss the causality and duality constraints
and find our picture is only the consistent way.

The identification of bulk operators correspond to CFT operators in a subregion is also important,
in particular for the understanding of the quantum entanglement in the AdS/CFT.
We identify these explicitly and show that the causal wedge naturally appears for the ball shaped region.
However, our result show that a (version of) subregion duality is not hold.
This subregion duality is partly based on the HKLL AdS-Rindler reconstruction,
then there are some misunderstandings of it.
Indeed, we show that bulk correlation functions in the
HKLL AdS-Rindler reconstruction and the corresponding ones in the HKLL global AdS reconstruction
are different 
%and the reconstructed bulk local operators are different for these two,
due to the null geodesics discussed in \cite{Bousso:2012mh}.
This means that the bulk local operators reconstructed from these two reconstructions are different even in the low energy.

Our picture of the bulk reconstruction can be applied to 
the asymptotic AdS spacetime, assuming the BDHM relation.
In the CFT, this background is a non-trivial state which has the bulk semi-classical description.
In particular, we argue that the bulk reconstruction for the (single sided) black hole is only possible
for the spacetime outside the (stretched) horizon.

This paper is organized as follows.
In the next section
we give the simple reconstruction of the bulk local operators and bulk wave packets.
We also study the bulk operators for CFT operators in a subregion
and discuss the subregion duality and the HKLL AdS-Rindler reconstruction.
In section three, 
we generalize our reconstruction picture to bulk theories on 
asymptotic AdS spacetimes.

\section{Simple reconstruction of bulk local operators 
}

\subsection{Relation between free scalar field on $AdS_{d+1}$ and large $N$ $CFT_d$}

In this subsection,
we will review the relation between free scalar field on $AdS_{d+1}$ and $CFT_d$ in the large $N$ limit
around the vacuum, 
where the bulk theory is free.

\begin{figure}
  \centering
  \includegraphics[width=4cm]{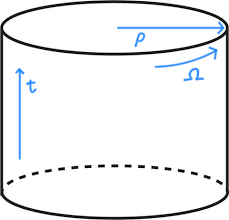}
  \caption{AdS space}
\label{fig1}
\end{figure}

Let us consider the free scalar field with the following action:
\begin{align} 
S_{scalar}= \int d^{d+1} x \sqrt{-\det (g)}
\left(
\frac12 g^{M N} \nabla_M \phi \nabla_N \phi
+ \frac{m^2}{2} \phi^2 
\right),
\end{align}
where $M,N=1, \cdots,d+1$,
on global $AdS_{d+1}$.\footnote{
In this paper, we consider the bulk scalar field only although the generalizations
of our picture to 
the gauge fields and the gravitons may be straightforward. 
}
The $AdS_{d+1}$ metric is
\begin{align}
d s^2_{AdS} = -(1+r^2) dt^2 +\frac{1}{1+r^2} d r^2+ r^2 d \Omega_{d-1}^2,
\end{align}
where $0 \leq  r < \infty$, $-\infty < t < \infty$
and $d \Omega_{d-1}^2$ is the metric for the $d-1$-dimensional 
round unit sphere $S^{d-1}$.
We set the AdS scale $l_{AdS}=1$ in this paper.
By the coordinate change $r=\tan \rho$,
the metric is also written as
\begin{align}
d s^2_{AdS} = \frac{1}{\cos^2 (\rho)} 
\left( 
-dt^2 +d \rho^2+ \sin^2 (\rho)  d \Omega_{d-1}^2 
\right),
\label{cood}
\end{align}
where 
$0 \leq  \rho < \pi/2$.
With this coordinate, the AdS spacetime is like a solid cylinder
as shown in Figure \ref{fig1}.
%With $z=\pi/2- \rho$, the metric is given by 
%\begin{align}
%d s^2_{AdS} = \frac{1}{\sin^2 (z)} 
%\left( 
%-dt^2 +d z^2+ \cos^2 (z)  d \Omega_{d-1}^2 
%\right).
%\end{align}
%The boundary of the $AdS_{d+1}$ is located at
%$\rho=\pi/2$ or $z=0$.
The CFT primary field which is dual to the bulk scalar is 
${\cal O}_\Delta (t,\Omega) $ with the conformal dimension $\Delta=d/2+\sqrt{m^2+d^2/4}$
on the cylinder $d s^2_{cyl} = 
-dt^2 +d \rho^2+ \sin^2 (\rho)  d \Omega_{d-1}^2$.\footnote{
In this paper, we consider this choice of $\Delta$, which implies $\Delta \geq d/2$, for simplicity.}

The BDHM relation \cite{BDHM},
\begin{align} 
\lim_{z \rightarrow 0} 
{\phi (t,z,\Omega)
\over z^\Delta}
\sim 
{\cal O}_\Delta (t,\Omega), \,\,\,
\mbox{    where $z=\pi/2-\rho$}
\label{BDHMrelation}
\end{align}
is the relation between the bulk field and the CFT scalar primary field.
%${\cal O}_\Delta (t,\Omega) $ with the conformal dimension $\Delta$.
Using this relation and the bulk equations of motion, 
the bulk local field is reconstructed from the CFT primary field, as in the HKLL bulk reconstruction \cite{HKLL},
in the large $N$ limit where the bulk theory is free.
Thus, in this limit,
we can explicitly study
the bulk local operators from the  CFT viewpoint.
For this study, it is more convenient to express the 
operators or the states in the energy eigen state basis.
This was done explicitly in \cite{op} and 
the relation between the  bulk local operators and 
the CFT primary fields is clear
as shown in the Appendix \ref{A}.

\subsection{Reconstruction of bulk local operators 
}

Below, using the HKLL bulk reconstruction formula \cite{HKLL},
we will see how the bulk local operator is reconstructed.
The results are rather simple.
For example,
the bulk local state at center
is given by time evolution
of uniformly distributed
CFT states which are localized at the fixed time.
%In this paper, CFT primary states means that  
%the CFT primary operators acting on the vacuum
%although usually they are used only for CFT primary operators at $t=-\infty$.
Thus, we conclude that
through this kind of analysis, we understand
how the bulk local operator emerges.

\begin{figure}
  \centering
  \includegraphics[width=8cm]{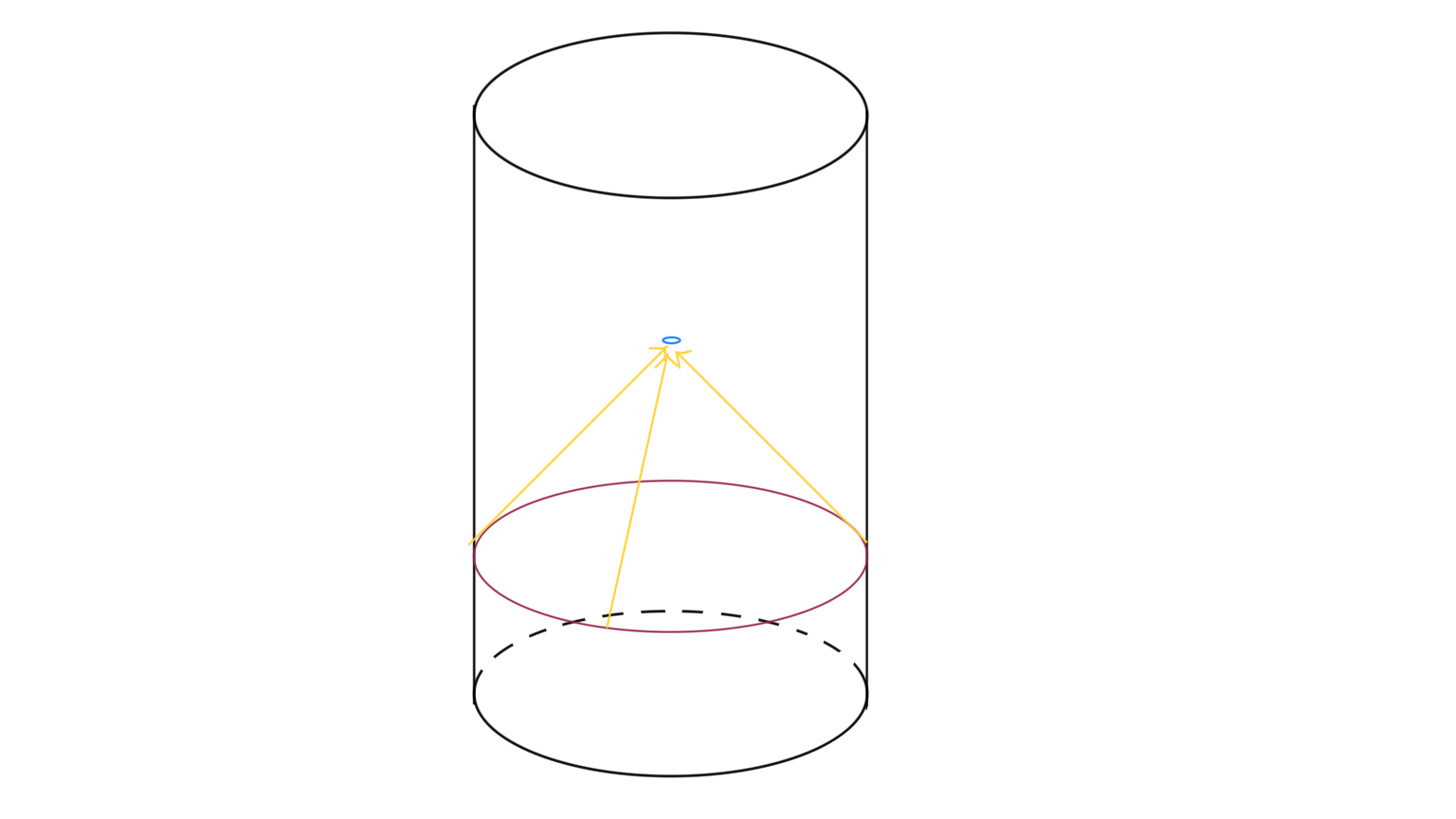}
  \caption{
The blue dot represents the center ($\rho=0$) at $t=0$ where we consider the bulk local state $\phi(\rho=0) \ket{0}$. 
The red circle represents the $t=-\pi /2 $ slice on the boundary where we consider  $\int_{S^{d-1}}  d \Omega \, {\cal O}_\Delta (\Omega) | 0 \rangle$, which is equivalent to the bulk local state by the time evolution.
The yellow lines represent the light-like trajectories from the $t=-\pi /2 $ slice on the boundary to the center at $t=0$.  }
\label{fig2}
\end{figure}

Let us consider the bulk local operator at the center of AdS space, i.e. $\phi(\rho=0)$, at $t=0$
and the corresponding bulk local state at the center, $\phi(\rho=0) \ket{0}$.
It was shown in \cite{local} that this bulk state is essentially equivalent to 
the following CFT state: $\int_{S^{d-1}}  d \Omega \, e^{i {\pi \over 2} H} {\cal O}_\Delta (\Omega) | 0 \rangle$, i.e. 
\begin{align} 
\phi(\rho=0) \ket{0}=
\int_{S^{d-1}} d \Omega \, e^{i {\pi \over 2} H} {\cal O}_\Delta (\Omega) | 0 \rangle,
\label{sc}
\end{align}
up to an overall numerical constant (Figure \ref{fig2}).
More precisely, this expression is valid only for $\Delta=d-1$ and 
%An explicit example of this equivalence 
the general expression, which includes the time-derivatives, 
is given in the Appendix \ref{A}.
This CFT state $\int_{S^{d-1}}  d \Omega \, e^{i {\pi \over 2} H} {\cal O}_\Delta (\Omega) | 0 \rangle$ 
at $t=0$ is obtained from the CFT state
 $\int_{S^{d-1}}  d \Omega \, {\cal O}_\Delta (\Omega) | 0 \rangle$
at $t=-\pi/2$ by the time evolution to $t=0$.\footnote{
Here, we considered the state. The corresponding operator is
 $ {\cal O}_\Delta (\Omega, t=-\pi/2) =
 (e^{- {\pi \over 2} \frac{\partial}{\partial t}} {\cal O}_\Delta (\Omega,t) )|_{t=0} $.
}
(The operator $\int_{S^{d-1}}  d \Omega \, {\cal O}_\Delta (\Omega)$
is a CFT local state averaged over the whole space
and then it is invariant under the rotation.)
%XXX Figure XX

This equivalence, or the bulk reconstruction, has a simple picture
if we regard the CFT local operators 
%\footnote{
%As explained in \cite{}, in the low energy theory we can use the 
%following  approximation:
%${\cal O}_\Delta (\Omega,t_0) \sim 
%\sum_{n=0}^{n_{\textrm{cut-off}}} \frac{(t_0)^n}{n!} \frac{\partial^n}{\partial t^n}  
%{\cal O}_\Delta (\Omega,t)|_{t=0}$ where $n_{\textrm{cut-off}} \gg 1$ is 
%related to the energy cut-off, for example, the plank scale.
%In this sense, it is regarded as the CFT local operator at $(\Omega_0,t=0)$.
%} 
as the bulk operators on the boundary using the BDHM relation,
%as described by the figure XXX.
The bulk local operator at the center is formed by 
the boundary operators located 
on $S^{d-1}$ at $t=-\pi/2$,
which are reached to the bulk local operator by the
(spherical symmetric) light-like rays.
%XXX Figure XX

We note that the bulk local operator contains arbitrary high momentum modes 
because it is supported on a point,
thus the mass of the bulk scalar field is negligible, which is responsible for the light-like behavior. \footnote{
The local state is a kind of a sum of very small wave packets, which have very high momentum, and then it behaves like a massless field. 
We will explain this more precisely. The local operator itself is not well-defined operator and we need a smearing of it with a length scale $1/M$, for example, by the Gaussian factor. Because this $M$ is like an energy cut-off, the number of modes effectively contained in this smeared operator is proportional to $M^{d}$. The mass is negligible for energy modes with high energy compared with the mass. 
If $M$ is %taken to be 
comparable with the mass of the scalar field, then mass can not be neglected. However, for the local operator, $M$ should be much larger than the mass, then the mass is negligible for almost all modes. 
If we consider the corresponding state, the low energy modes are suppressed for large $M$ because of the normalization of the state which contains huge number of the modes. Therefore, the effect of the mass is suppressed by the cut-off $1/M$. 
One might think that the two point function of the local operator depends on the mass, which contradicts with the above statement. However, it is well-known that the two point function (retarded or Feynman) is divergent if the two operator insertion points are connected by a null-geodesics, even for the massive scalar field. This divergence is regularized by $M$ and the mass is negligible in the large $M$ limit. Thus, there is no contradiction. Of course, if $M$ is not taken to be large compared with the mass, the result depends on the mass.
}

\begin{figure}
  \centering
  \includegraphics[width=8cm]{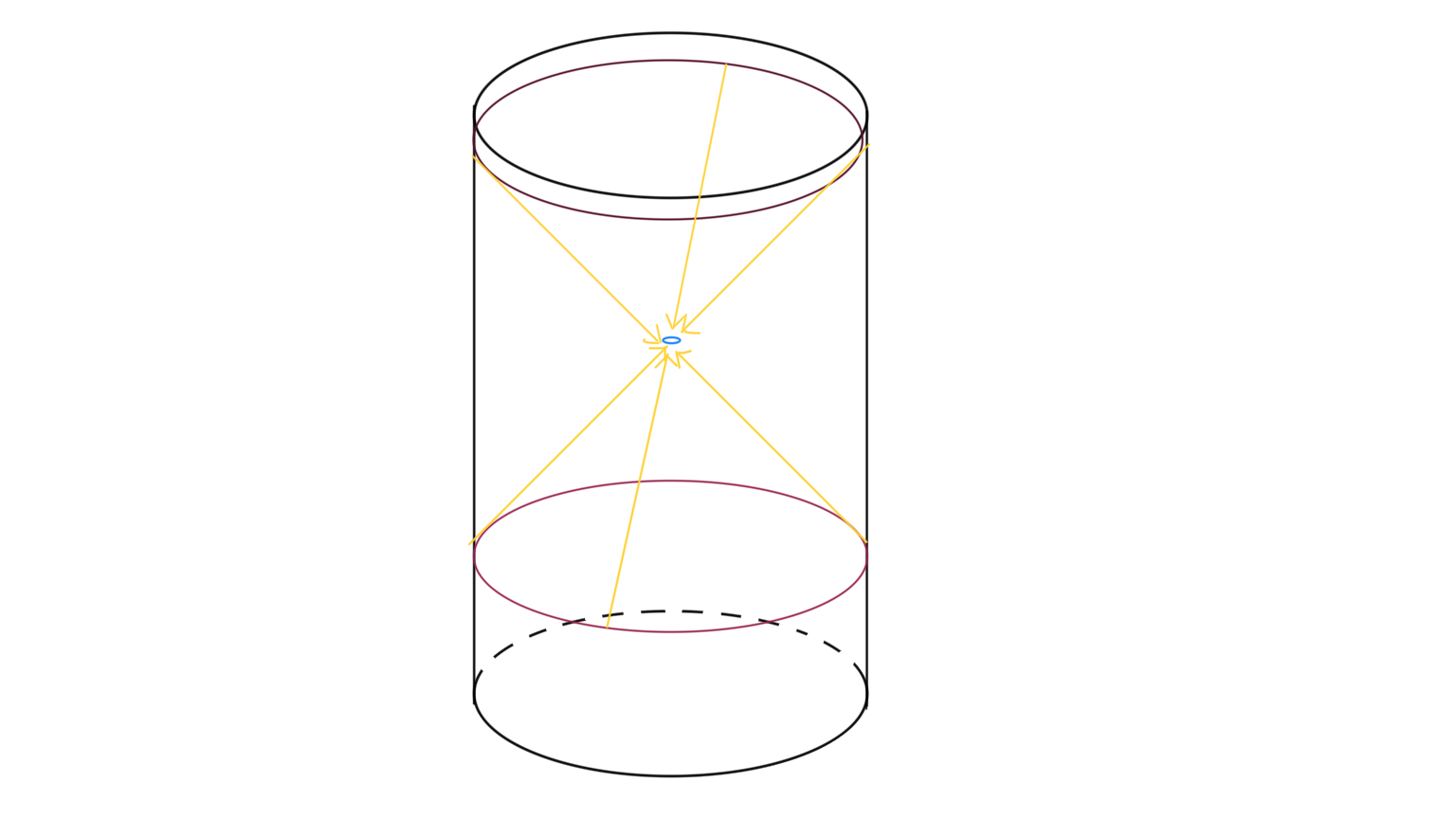}
  \caption{The bulk reconstruction by \eqref{formal}. Instead of the interval $-\pi/2 \geq t \geq \pi/2$, only the contributions from the $t= \pm \pi/2$ slices are needed to reconstruct the bulk local operator. }
\label{fig25}
\end{figure}

One might think this is different from 
the well-known HKLL bulk reconstruction \cite{HKLL}, in which 
the bulk local operator is constructed from the CFT operators 
on the time-interval $-\frac{\pi}{2} \leq t \leq \frac{\pi}{2} $,
not just on $t=-\pi/2$.
Below, we will show that the HKLL bulk reconstruction indeed give the 
same picture by a careful analysis.
Of course, this is not surprising because the HKLL bulk reconstruction and 
the bulk reconstruction given in Appendix \ref{A} are equivalent. 
The HKLL bulk reconstruction of the bulk local operator at the center is the following: 
\begin{align} 
\phi(\rho=0,t=0)
=\int_{-\frac{\pi}{2}}^{  \frac{\pi}{2} } d t' 
\int_{S^{d-1}}  d \Omega' \, K( \Omega' , t') {\cal O}_{\Delta} (\Omega', t'),
\label{HKLL eq}
\end{align}
where the smearing function $K( \Omega' , t') $ for $d=\mbox{odd}$ is given by
\begin{align} 
K(\Omega , t)  \sim 
{1 \over (\cos t)^{d-\Delta} }
\end{align}
up to a constant depending on $\Delta$ and $ d$.
For $d=\mbox{even}$, the smearing function is 
$K(\Omega , t)  \sim 
{\log t \over (\cos t)^{d-\Delta} }$. 
If $d-\Delta \geq 1$, 
an integration $\int d t'  K( \Omega' , t')$ is divergent 
near $|t'|=\pi/2$, which means that 
the contributions of the CFT local operator $ {\cal O}_{\Delta} (\Omega', t')$
in the HKLL reconstruction  (\ref{HKLL eq})
is negligible for $|t'|<\pi/2$.\footnote{
The divergence will be related to the divergence of the local operators
as the operators acting on the Hilbert space and we need some regularization as 
discussed in  \cite{local}. 
%For $\Delta=d-1$, the expression \eqref{formal} itself is correct.
%For $\Delta \neq d-1$, the correct expression with the CFT operators on $t=\pm \pi/2$
%is shown in Appendix \ref{A}.
}
Thus the time integration is localized on $t'=\pm \pi/2$:
\begin{align} 
\phi(\rho=0,t=0)
\sim 
\int_{S^{d-1}}  d \Omega' \,  
\left(
{\cal O}_{\Delta} (\Omega', t=-\pi/2)
+{\cal O}_{\Delta} (\Omega', t=\pi/2)
\right),
\label{formal}
\end{align}
as shown in Figure \ref{fig25}.
(Again, this expression is valid only for $\Delta=d-1$ and 
the general expression, which includes the time-derivatives, 
is given in the Appendix \ref{A}.)
The previous picture is obtained for the bulk local state $\phi(\rho=0,t=0) \ket{0}$
if we notice that 
the CFT operator $\int_{S^{d-1}}  d \Omega' \,  {\cal O}^+_{\Delta} (\Omega', t)$,
where $ {\cal O}^+_{\Delta} (\Omega', t)$ is the positive energy modes of $ {\cal O}_{\Delta} (\Omega', t)$, at $t=\pi/2$ is same as the one at $t=-\pi/2$ up to a overall constant because of the time periodicity in the large $N$ limit and the parity invariance.

Note that this localization of the integration is indeed required by the bulk causality
and the BDHM relation $ {\cal O}_{\Delta} (\Omega, t) \sim \phi(\rho=\pi/2,\Omega,t) $,
i.e. the identification of the CFT primary operator as the bulk operator at the boundary.
The bulk operator at the center $\phi(\rho=0,t=0)$
%$\phi(\rho=\pi/2,\Omega,t)$ 
should be independent from the bulk operator at the spacetime point outside the 
light cone of the center $\{ \rho=0,t=0 \}$
%the boundary point $(\rho=\pi/2,\Omega,t)$, 
however,
the boundary point $\{ \rho=\pi/2,\Omega,t \}$ is obviously not within the light cone for $|t|<\pi/2$.
Thus, the bulk reconstruction of $\phi(\rho=0,t=0)$ should commute with
any local operator constructed from $ {\cal O}_{\Delta} (\Omega,t)$ for $|t|<\pi/2$.
Thus, the CFT operators on $ {\cal O}_{\Delta} (\Omega,t)$ for $|t| =\pi/2$
should be dominant for the reconstruction of the bulk local operator and
our picture is natural and consistent with the bulk causality. %\footnote{
%On the other hand, the time like separated point from the center at $t=0$ does not appear
%because the bulk local operator has an infinite energy formally \cite{local}.}
In this sense, we can say that the HKLL reconstruction formula \eqref{HKLL eq} is correct, but misleading.

For $d-\Delta < 1$, 
the HKLL reconstruction seems to give a completely different picture
because the smearing function is zero on $t=\pm \pi/2$. 
However, even for this case, the above picture, in which only the CFT operators on 
an arbitrary small region containing
$t=\pm \pi/2$ are relevant, is correct as we will see in appendix \ref{A2},
essentially because $t=\pm \pi/2$ is the branch point, thus the singular point,  of the hypergeometric 
function appeared in the HKLL reconstruction.

Next, we will consider the bulk local operator which is not at the center, 
$\phi(\rho, \Omega,t=0)$, and the CFT operator corresponding to it.
For this, the HKLL reconstruction formula was obtained by the conformal map from 
the formula for the center \cite{HKLL}.
Thus, the above picture is also true for this case.
More explicitly, the reconstruction formula is same as (\ref{HKLL eq}),
\begin{align} 
\phi(\rho, \Omega,t=0)
=\int d t' 
\int d \Omega' \, K( \rho, \Omega, \Omega' , t') {\cal O}_{\Delta} (\Omega', t'),
\end{align}
where the integrations of $t'$ and $ \Omega'$ were taken over the  
boundary points which are space-like separated from the bulk operator insertion point $\{ \rho, \Omega, t=0 \}$ and
the smearing function is 
\begin{align} 
K(\rho, \Omega, \Omega' , t')  \sim 
{1 \over (d(\rho,\Omega, \Omega',t'))^{d-\Delta} },
\end{align}
where $d(\rho,\Omega, \Omega',t')$ is the geodesic distance between the point $\{ \rho, \Omega, t=0 \}$ and the boundary point $\{ \rho=\pi/2, \Omega', t' \}$.
The geodesic distances are zero for the the boundary points connected to the 
bulk operator insertion point by the light-like trajectories,
then the integration is localized on such boundary points.
This means that 
the bulk local operator is composed by the CFT primary operators 
on such boundary points\footnote{
The integration is localized on 
the future and the past boundary points.
If we consider the $\phi^+$ or the corresponding state, the integrations on these two regions is same up to a constant as for case with the bulk local operator at the center.
This is because the time translation $t \rightarrow t+\pi$ gives the factor $e^{i \pi (2n+l+\Delta)}$ 
and the parity transformation in space $S^{d-1}$ gives $(-1)^l$ \cite{local}.
The combination of these gives only a phase $e^{i \pi \Delta}$. 
}
and the bulk reconstruction picture above is true for this case also.\footnote{
We have considered the one particle state.
For multi-particle states, we have 
$ \phi(\rho_1, \Omega_1  ,t=0) \phi(\rho_2, \Omega_2  ,t=0) \cdots \phi(\rho_q, \Omega_q  ,t=0) \ket{0} = 
\phi^+(\rho_1, \Omega_1  ,t=0) \phi^+(\rho_2, \Omega_2  ,t=0) \cdots \phi^+(\rho_q, \Omega_q  ,t=0) \ket{0}$  
in the large $N$ limit if  $\{ \rho_i, \Omega_i \}$ are different each other.
Thus, the bulk reconstruction of the state can be done from the CFT operators on only the past boundary points.
}

\subsection{Bulk wave packet moving in a particular direction}

We have seen that a bulk local state is represented by 
CFT primary states integrated on 
a space-like surface, which is regarded as a time-slice with an appropriate definition of a time.
In this section we will consider CFT primary operators integrated on a small and an almost point like region, 
instead of the space-like surface.
We will see that the bulk local state moving in a particular direction\footnote{
More precisely, what we will consider is a kind of wave packets 
whose size is much larger than the cut-off (Plank) scale, but much smaller than
any other length scale.
}
corresponds to a state obtained by acting such a CFT operator on the vacuum.

We will consider the following operator:
\begin{align} 
\phi_A \equiv
\int_{S^{d-1}} d \Omega \, f_A(\Omega) \,e^{i {\pi \over 2} H}   {\cal O}_\Delta (\Omega) e^{-i {\pi \over 2} H} ,
\label{wac}
\end{align}
where 
$f_A(\Omega)$ is a function (effectively) supported 
on a small region $A$ in $S^{d-1}$.
More explicitly, we take a small ball shaped region with radius $l_A$ for $A$ 
and  the Gaussian distribution for $f_A(\Omega)$.
If we averaged the position of $A$ over the whole space $S^{d-1}$, 
this operator becomes the bulk local operator at the center, as we have seen.
Precisely speaking, we need to act some derivatives on $ {\cal O}_\Delta (\Omega)$ in \eqref{wac}
%if $\Delta \neq d/2-1/2$ 
as explained in Appendix \ref{A1}. 
Below, we will omit such derivatives for the notational simplicity.

Thus, it is natural to think $\phi_A$ represents a part of the bulk local operator
at the center. 
Indeed, in \cite{local}, it was shown that the 
$\phi_A \ket{0}$ is a bulk (almost) local state or a wave packet at the center moving in 
the radial direction.
Note that the bulk local operator should be smeared over a small, but larger than 
the Plank scale region. 

\begin{figure}
  \centering
  \includegraphics[width=8cm]{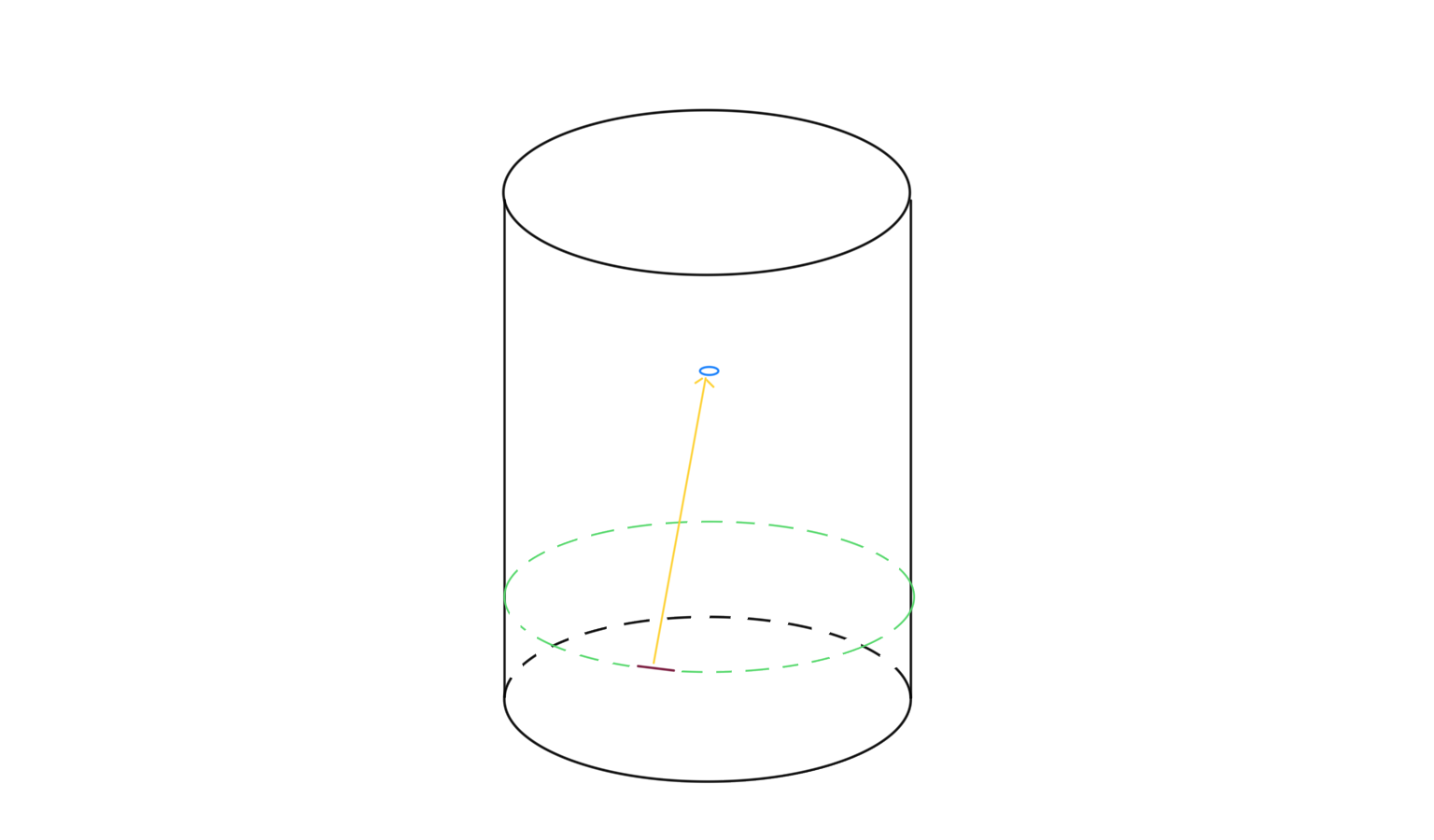}
  \caption{The bulk wave packet moving toward the center.
The small region $A$, which is represented by small red curves, is on 
the $t=-\pi/2 $ slice on the boundary, which is represented by 
the green dotted circle.
The yellow line to the center is the light-like trajectory from $A$ 
and represents the wave packet.}
\label{fig3}
\end{figure}

Furthermore, 
the following operator:
\begin{align} 
\phi_A(z) \equiv
\int_{S^{d-1}} d \Omega \, f_A(\Omega) \,e^{i {z} H}   {\cal O}_\Delta (\Omega) e^{-i {z} H},
\end{align}
was shown to be a bulk wave packet moving inward  in 
the radial direction at $\rho=\pi/2-z$, where $\pi/2 \geq z>0$,
and $\Omega$ is the location of the small region $A$,\footnote{
For $\pi/2 \leq z <0$, the corresponding bulk wave packet
is at $\rho=\pi/2-|z|$ and moving outward in the radial direction.
}
 \cite{local}.
Of course, this CFT operator is the time evolution of the 
CFT operator $\int_{S^{d-1}} d \Omega \, f_A(\Omega) \, {\cal O}_\Delta (\Omega)$ at $t=-z$ to $t=0$. (Figure \ref{fig3}.)

This result also has a simple interpretation in the bulk picture as before. 
The CFT operator of $\phi_A$ is regarded as 
the bulk operator on the boundary smeared over
the region $A$ by the Gaussian distribution, but not smeared for the radial direction. 
(More precisely, it should be smeared for the time-direction with some length scale which is
very small, but larger than the Plank scale \cite{local}.)
This is a kind of bulk wave packets.
A usual wave packet is defined by choosing a particular momentum.
For our case, the momentum for the angular direction is  ${\cal O} (1/l_A)$,
but the momentum for the radial direction is not constrained, then
most of them are much larger than $1/l_A$.
Thus, the operator is a linear combination of wave packets
and most of them are moving in the radial direction.
This means that at $t=0$ the wave packet is at $\rho=-z$ 
%(and at the center for $z=\pi/2$)
and 
the $\phi_A(z)$ represents the small wave packet localized at the center
moving in the radial direction.

\begin{figure}
  \centering
  \includegraphics[width=8cm]{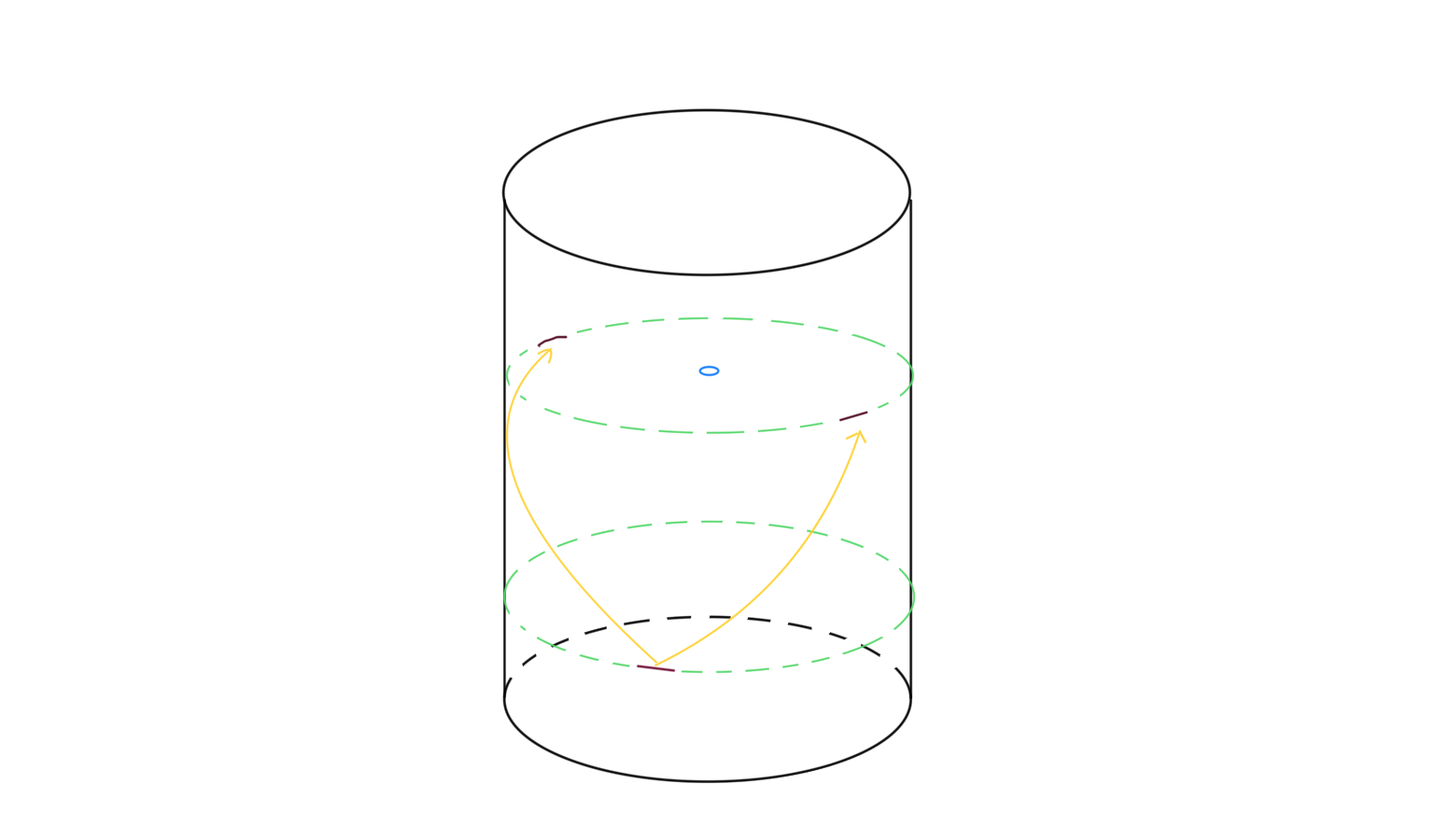}
  \caption{The CFT picture of Figure \ref{fig3}.
Here, the two green dotted circles represent the $t=0 $ and $t=-\pi/2 $ slices.
The two yellow curves on the boundary from $A$ to the antipodal points on the boundary 
are the light-like trajectories from $A$ 
and represent the wave packet in the CFT picture.
 }
\label{fig4}
\end{figure}

On the other hand, 
in the CFT picture, 
the operator localized in the region $A$ at $t=-\pi/2$
is spread out spherically $(S^{d-2})$ in space ($S^{d-1}$) at $t=0$.
Here, very large momentum modes up to $1/l_A$ are contained in the operator
and this spreading is light-like \cite{comm}.\footnote{
The non-trivial CFT commutator  
in the K\"{a}ll\'{e}n-Lehmann like representation contains the infinitely many massive modes \cite{comm} and one might think that this light-like spreading is not valid.
In the Appendix \ref{B1}, we will show that this is indeed valid. 
}
Thus, the operator at $t=0$ is localized in a "great circle" in $S^{d-1}$.
For the $d=2$ case, the "great circle" is two antipodal points in $S^1$.
(Figure \ref{fig4}.)
Even for $d>2$, 
if we take $A$ in the definition of $\phi_A$ 
as an ellipsoid which is squeezed in a particular direction,
then the operator $\phi_A$ is 
localized on the two antipodal points in $S^{d-1}$ in the CFT picture.
Furthermore, this $\phi_A$ is still localized at the center 
in the bulk picture if we take the length scales of the ellipsoid much larger
than the cut-off scale (smearing) of the local operator
as like the ball shaped region case.

This implies that this bulk localized wave packet is constructed as
the entangled state in CFT, which can be written schematically as
$(\ket{1} \otimes \ket{0}+\ket{0} \otimes \ket{1}) \otimes \ket{\rm others}$,
where $\ket{1}$ is the one particle state, $\ket{0}$ is no excitation state 
and, the first and the second kets represent the states localized on the two antipodal points.
On the other hand, %it should be noted that 
a linear combination of CFT primary states (with a small number of derivatives) at two different points is a linear combination of bulk states
localized on two points on the boundary, not inside, even in the bulk picture,
although this is also an entangled state.
\footnote{
\label{foot12}
This implies that if we choose one of the antipodal points at $t=0$ in Figure \ref{fig4} and
extract the state localized there,
it can not be realized by the primary operators (with a small number of derivatives).
The primary operators with a large number of derivatives on the antipodal points at $t=0$ 
can realize this state, which is the local state there.
However, this is effectively a non-local state in the low energy theory because of the derivatives, 
which can be regarded as a result of the Reeh-Schlieder theorem \cite{local}.
%Furthermore, it can not represent a bulk operator near the center because 
%it is not reflection symmetric. 
}

We have seen that the bulk local operator moving in the radial direction
is represented by $\phi_A(z)$ in the CFT picture.
It is easy to generalize this to the bulk local operator moving in 
an arbitrary direction by considering a conformal map,
which maps the bulk local operator at the center to an arbitrary point,
say $\rho=\rho_1, \Omega=\Omega_1, t=t_1$ \cite{HKLL}.
%(We made an time-translation such that the point is on $t=0$.)
This conformal map changes the $t=-\pi/2$ slice $(S^{d-1})$ on the boundary
to the slice on $t=-\pi/2+\delta t(\Omega)$ where $\delta t(\Omega)$ represents
the tilt in time direction.
Thus, if we 
generalize $\phi_A(z)$ to
\begin{align} 
\phi_A(z)  & \equiv
\int_{S^{d-1}} d \Omega \, f_A(\Omega) \,e^{i {z} H}   {\cal O}_\Delta (\Omega,t=\delta t(\Omega)) e^{-i {z} H} \\
&= \int_{S^{d-1}} d \Omega \, f_A(\Omega) \,e^{i {(z+\delta t(\Omega))} H}   {\cal O}_\Delta (\Omega) e^{-i {(z+\delta t(\Omega))} H},
\label{phiA}
\end{align}
this $\phi_A(z)$ represents the bulk local operator which 
moving from the small region $A$ on $t=-\pi/2$ slice to the point 
$\rho=\rho_1, \Omega=\Omega_1, t=t_1$ light-likely
where $z$ parametrizes the light-like trajectory. 
The wave packet is at the the small region $A$ for $z=0$ and at the bulk point $\rho=\rho_1, \Omega=\Omega_1, t=t_1$ for $z=\pi/2+t_1$.
This picture is also checked in the operator formalism as we will see 
in the Appendix \ref{A3}.

In summary, we have the following simple and intuitive picture of 
the reconstruction of the bulk local operator or state in AdS/CFT around the vacuum.
The bulk local state at a spacetime point $P$ 
is represented by
the CFT primary state integrated over $C$,
where the region $C$ is the intersection of  
the light rays emanating from $P$ to the past
and the boundary of AdS spacetime.
The wave packet%\footnote{
%The wave packet here is defined by smearing of the bulk local operator
%for all directions
%except a direction for which the (most of) momenta take values.}
 of the bulk local state at the point $P$
is represented by
the CFT primary state integrated over $A$,
where very small region $A$ is the intersection of  trajectory of the wave packet
to the past
and the boundary of AdS spacetime.
For the bulk local operator instead of the state,
we just need to take into account the light rays to the future adding to ones to the past .

\subsubsection{ Causalities and duality}

Here, we will check that the above picture is consistent with 
the causalities and the duality.
For the light-like trajectory from a boundary point toward
the center, as drawn in Figure \ref{fig3} and Figure \ref{fig4}, 
it was already shown in \cite{local} that the bulk causality is consistent 
with the CFT picture.
Indeed, both in the CFT and bulk pictures,
the (light-like) wave packets reach the boundary at the antipodal point 
at the same time.
This result was already shown in \cite{Gao-Wald}
where, in AdS spacetime,
any light-like trajectory from a boundary point at $t=0$ 
reaches the antipodal point on the boundary at $t=\pi$.
Such light-like trajectory can be always on the boundary,
which is realized in the CFT picture.
Thus, any light-like trajectory from a boundary point in the bulk picture
and the corresponding two light-like trajectories in our CFT picture 
reach the antipodal boundary point at same time, then this picture is 
consistent with the causality.
Furthermore this is the only possibility for the consistent CFT dual to 
the bulk wave packet\footnote{
In this paper, we take $\Delta={\cal O}(1)={\cal O}(1/l_{AdS})$ .
This implies that if we consider the wave packet whose size should be much smaller
than the AdS scale $l_{AdS}$, the mass is negligible and the trajectory 
of the wave packet is light-like.
}
if we require the causalities both in the bulk and the CFT pictures.\footnote{
The causality constraints were also studied in \cite{Berenstein} although 
the bulk reconstruction picture is different from ours.}
This is because the bulk wave packet on the boundary,
which is realized when $t=0$ or $t=\pi$, is regarded
as a wave packet composed by the CFT primary field
and the speed of the propagation is bounded by the speed of light
in the both pictures.
This implies that in the CFT picture also, the corresponding wave packets should be 
on the light-like trajectories.

\subsubsection{ Locality in radial direction }

Here, we will discuss how the locality in radial direction 
appears in our picture of the bulk reconstruction. 
Note that the discussions here are only for the free theory limit in the bulk.
For the spherical symmetric states (e.g. in Figure \ref{fig2}), 
the bulk local state integrated over the sphere $S^{d-1}$ 
at the radial coordinate $\rho=\pi/2-z_0$ 
corresponds to the CFT primary state on the time $t=-z_0$ slice integrated over the sphere $S^{d-1}$.
This implies that the radial locality is same as the locality in the time direction
in this setup.
In order to realize this, it is needed that the primary fields
with different times are (almost) independent (for $-\pi/2< t < \pi/2$).
This is only possible for the theory with 
(almost) infinite degrees of freedoms like the large $N$ gauge theory.
Indeed, for the free CFT or a theory with a finite degrees of freedom,
the fields with different time are related by the equation of motions
and they are not independent.

For the non-spherical symmetric case,
the radial location of the bulk local operator is 
related to the time coordinate of the CFT primary fields
although the relation is not direct.
The independence between CFT primary fields at different times 
are important for this case also.
This independence means that commutator of the 
CFT primary fields at different times vanish.
For the free CFT, the commutators of two free fields does not vanish
if the one field is in the light-cone of the other field.
Of course, this property is same for any CFT, however,
for the non-trivial CFT, a commutator of the 
CFT primary fields is non-zero value on the light-cone and 
is diverging at $\sin t=0$ where $t$ is the time difference between the two fields, as was explicitly shown for $\Delta=d/2$ \cite{comm}.
This divergence is regularized by the smearing of the local operator
with a cut-off
and the normalized commutator by this cut-off is effectively zero 
for CFT primary fields at different times.

Note that this discussion can be applied for any non-trivial CFT.
The holographic CFT\footnote{
In the operator formalism, 
the holographic CFT is the CFT with the properties assumed in \cite{op}.
} 
is special because 
it behaves like the generalized free field,
which implies that the product of the CFT primary fields 
are independent each other.
In the commutator or the OPE languages, 
this is reflected in the fact that the contribution proportional to the identity operator
is dominant in the large $N$ limit.
Thus, in our picture, the bulk locality in radial direction is essentially
mapped to the locality in the time direction in the CFT.

\subsection{Bulk operators correspond to  CFT operators in a region}

We have seen how bulk local operators are mapped to 
the CFT operators.
We can also map
the CFT operators supported in a region $A$ in $S^{d-1}$ on $t=0$ slice, $\{ {\cal O}_A \}$,
to the bulk operators on $t=0$ slice \cite{local}.
In particular, we will consider 
what is the (smallest) bulk region $a$ corresponding to the CFT region $A$,
where the bulk operators corresponding to $\{ {\cal O}_A \}$ 
are supported in the bulk region $a$.
(Note that if the subregion duality \cite{Wa}\cite{Bo} is correct, 
%the bulk operators supported in a region $a$
%should be equivalent to the CFT operators supported in a region $A$ 
these bulk operators should be equivalent to $\{ {\cal O}_A \}$, i.e. 
any bulk operator supported in the region $a$ should correspond to some ${\cal O}_A$,
in the low energy approximation. 
Here we do not assume this version of the duality.
We will call it the strong version of the subregion duality, in order to distinguish it
to a version of the subregion duality where the correspondence between the density matrices is required,
as we will explain later.)

We note that any CFT operator %supported in space $S^{d-1}$ 
(at $t=0$)
can be generated\footnote{
In the large $N$ limit, the bulk theory is free 
i.e. the CFT we consider is the generalized free theory.
Thus, properties of a product of $\phi_f$'s, which correspond to multiple particles, follows from
the properties of $\phi_f$, which corresponds to one particle.
}
by the CFT primary operator integrated over 
%the whole space and an interval of time
the spacetime\footnote{
Of course, here we regard
 $ {\cal O}_\Delta (\Omega,t)=e^{i t H}  {\cal O}_\Delta (\Omega) e^{-i t H} $ as an operator at $t=0$ slice.}
with a weight $f(\Omega,t)$:
\begin{align} 
\phi_{f} \equiv
%\int_{-\pi}^{\pi} dt
\int  dt
\int_{S^{d-1}} d \Omega \, f(\Omega,t)   {\cal O}_\Delta (\Omega,t),
\end{align}
because derivatives of the primary field can be represented 
by taking the distribution $f(\Omega,t)$ as derivatives of the (smeared) delta function. 
%Note that the time translation and the parity symmetries enable us to
%restrict the integration of time to the interval.

\begin{figure}
  \centering
  \includegraphics[width=5cm]{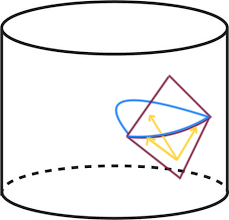}
  \caption{The red lines represent the region $A$ on $t=0$ slice and 
its causal diamond on the boundary. 
The causal wedge of $A$ on $t=0$ slice is the bulk region inside of the blue curve.
The yellow lines are light-like trajectories. 
The one of the these lines toward the inside the bulk represents the wave packet in the causal wedge.
The other two are on the boundary and represent the wave packets in the CFT picture.
}
\label{fig5}
\end{figure}

Let us investigate which $\phi_f$ is a CFT operator 
supported in the region $A$.
It is obvious that 
$\phi_f$ is supported in the region $A$
if $f(\Omega,t)$ is %only supported
nonzero only 
in the causal diamond of the region $A$ in the CFT picture
because of the causality (Figure \ref{fig5}).
More generally, as we have seen in the previous section,
a small wave packet of CFT local operators at $(\Omega_0,t_0)$ 
can behave like two light-rays emitted from there to opposite directions in space $S^{d-1}$.
If both of the two light rays reach the region $A$ on $t=0$ slice,
this wave packet of CFT local operators is supported in the region $A$.
Thus, by taking $f(\Omega,t) $ as this wave packet at $(\Omega_0,t_0)$,
$\phi_f$ is an operator supported in the region $A$.
%Note that a wave packet moving in one direction created by the CFT primary field corresponds to a bulk local operator
%on the boundary and if it reachs a point in the region $A$ on $t=0$, then it also can be represented by the CFT primary operator
%at the point on $t=0$.
It is difficult to expect other possibilities of choice of $f(\Omega,t)$
to obtain $\phi_f$ is an operator supported in the region $A$
because the propagation is generic other than the causality constraints.
Thus, below we assume that 
$\phi_f$ is a CFT operator 
supported on the region $A$ only if $f(\Omega,t)$ is a  linear combination
of such wave packets.

\subsubsection{Ball shaped region }

First, we take the subregion $A$ as a ball shaped region in $S^{d-1}$.
For this case, in the CFT picture, 
the above wave packets are in the causal diamond if 
the two light rays reach the region $A$ on $t=0$.
Thus, $\phi_f$ is supported on the region $A$
only if $f(\Omega,t)$ is nonzero only 
in the causal diamond of the region $A$  (Figure \ref{fig5}).

In the bulk picture, 
as we have seen in the previous section,
the bulk operator corresponding to such $\phi_f$ 
represents a linear combination of  
wave packets moving in arbitrary directions in bulk space.
Such wave packets in the bulk can reach only the causal wedge of $A$ 
on $t=0$.
This means that 
any CFT operator 
supported on the region $A$ corresponds 
to a bulk operator supported on the causal wedge of $A$.
Note that the causal wedge (on $t=0$) is the bulk region inside
the Ryu-Takayanagi surface of $A$ \cite{RT}, i.e. the entanglement wedge.\footnote{
The entanglement entropy for a region $a$
in a QFT is known to be proportional to the area of the region $a$.
Thus, if we regard the perturbative bulk theory as a QFT,
Ryu-Takayanagi surface appears in our set up  
might be natural.
However, this area law depends on the UV cut-off.
Furthermore, for example, for $d=2$ CFT, the entanglement entropy depends
on the central charge $c$ only in the large $c$ limit although the bulk theory with a fixed central charge can have any number of 
the scalar fields which give different entanglement entropies as a QFT.
Thus, it is not obvious that the relation between 
the entanglement entropy computation
using the Ryu-Takayanagi surface
and the low energy bulk states correspond to the CFT region $A$.
}

\subsubsection{Subregion duality, error correction code and AdS-Rindler reconstruction}

It is important that there exist bulk operators 
supported in the causal wedge of $A$ which can not be reconstructed from
CFT operators
supported in the region $A$.
Indeed, there are bulk light-rays or null-geodesics starting from a boundary point to
another boundary point through the causal wedge of $A$ 
such that both of the two boundary points are not 
the (CFT) causal diamond of $A$.
Then, the wave packets along these can not reconstructed from 
CFT operators
supported in the region $A$.
Note that in the large $N$ (free bulk theory) limit we considered in the paper, any state with a (Plank scale) cut-off of the energy, which is realized by a smearing of the local operator, can be regarded as a low energy state, and states of the bulk free theory and the CFT (which is approximated as the generalized free theory) are identical. In other words, we restricted only the low energy states in this paper.
Thus, there is no alternative reconstruction from 
CFT operators
supported in the region $A$.

This has implications for the subregion duality \cite{Wa}-\cite{Bo}.
Indeed, this violates a strong version of the subregion duality,
which claims that 
any bulk operator 
supported in the causal wedge of $A$
can be reconstructed from 
the low energy CFT operators
supported in the region $A$,
for this set-up. 
(We consider only the low energy CFT operators in this paper.)
Problems of the strong version of subregion duality
related to such null-geodesics were already raised in \cite{Bousso:2012mh}.
In \cite{Bousso:2012mh}, it was stated that some non-local operators could solve the problems
although there are no concrete arguments for this.
In our setup  there are no such operators solving the problem because the low energy CFT spectrum
is given and the operators associated to the null-geodesics are explicitly given by 
the CFT primary operators, which are not supported in the region $A$.\footnote{
The precursor was introduced in \cite{Polchinski:1999yd} and this might be such operators.
However, the energy of the wave packet considered in \cite{Polchinski:1999yd} is string scale, which is infinite in the approximation in this paper.
Moreover, in our set up, we consider the state without introducing the source term
and the bulk local state at the center at some time, which is bouncing by the boundary periodically by the time evolution,
corresponds to the (non-local)  CFT state (\ref{sc}) which is different from the vacuum any time.
Thus, there is no need to introduce the precursor.
%In other words, the CFT operator creating this CFT state is the precursor.
}
Thus, the strong version of the subregion duality is not valid.\footnote{
In other words, 
$ {\cal A}_A^{CFT} \subset  {\cal A}_a^{bulk}$,
instead of the subregion duality
$ {\cal A}_A^{CFT} =  {\cal A}_a^{bulk}$,
in the low energy approximation,
where ${\cal A}_A^{CFT} $ is the set of the CFT operators supported on the ball shaped region $A$
and ${\cal A}_a^{bulk}$ is the set of the bulk operators supported on the causal wedge $a$ for the region $A$.
}
Note that this duality is assumed in many papers, for example,  in \cite{Dong:2016eik} for the entanglement wedge reconstruction. 

The strongest reason to believe 
this strong version of the subregion duality may be the
HKLL AdS-Rindler reconstruction \cite{HKLL}, by which 
the correlation function of the bulk local operators in the AdS-Rindler patch is reproduced from
certain CFT operators in the corresponding subregion $A$ \cite{Morrison}.
This seems to implies that the bulk local operators can be reconstructed from 
CFT operators in a subregion, not in the whole space $S^{d-1}$ as we have seen.
However, the HKLL AdS-Rindler reconstruction assume 
that the BDHM relation holds even in the AdS-Rindler patch.
This assumption was recently shown to be violated
 because of the finite $N$ effect \cite{Sugishita}, 
and the HKLL AdS-Rindler reconstruction is not correct.
Thus, 
 $\phi(x) \neq \phi_{\rm Rindler}(x) $,
where we denote by
 $ \phi_{\rm Rindler}(x) $ the part of the bulk local operator  
 $\phi(x)$ which can be reconstructed from the CFT operators supported on the Rindler subregion $A$ in the boundary.\footnote{
In the CFT picture, the modes of the region $A$ and the region $\bar{A}$ will be complete
as for the massless scalar in the Minkowski space.
(It is complete if the primary operators with a large number of derivatives 
are regarded as a local operators. In the low energy theory, such operators are not
regarded as a local operators \cite{local}
as mentioned in footnote \ref{foot12}.)
One might think that 
this means that 
the modes of the corresponding two causal wedges of $A$ and $\bar{A}$ are complete.
This is not correct because the reconstruction of the bulk local operators is non-local in the CFT picture
(and the strong version of the subregion duality is not valid).
Indeed, there exist the wave packets, which are almost bulk local operators, reconstructed from the CFT operators supported on both of the regions $A$ and $\bar{A}$, for the null-geodesics.  
Note also that
the bulk local operator itself is always supported on the whole space $S^{d-1}$ in the CFT picture, as seen in Figure \ref{fig2}. 
}
Note that the part of 
 $\phi(x) $ which can not be reconstructed from CFT on $A$ is related to the horizon to horizon null-geodesics in \cite{Sugishita}.

Note that the quantum error correction code proposal \cite{ADH} is partly based on
the claim that $\phi(x) = \phi_{\rm Rindler}(x) $ in the low energy theory,\footnote{
In this paper, we always consider the low energy theory and this equation 
is understood in the low energy theory.
} thus this proposal is not realized in holographic CFTs.
Even from the following discussion using the HKLL bulk reconstruction \eqref{HKLL eq}, it is easy to see that 
 $\phi(x) = \phi_{\rm Rindler}(x) $ 
is not valid.
By inserting \eqref{CFTop} into \eqref{HKLL eq} and integrating over the 
$S^{d-1}$ first, we can see that
the bulk local operator at the center $\phi(\rho=0) $ only contains the modes with
$l=0$, i.e. the spherical symmetric modes $a_{n00}$.
These modes are generically low energy modes. % if the radial quantum number $n$ is not large enough. 
On the other hand, it is obviously impossible to construct the operator, which only
contains the spherical symmetric modes, from
the CFT operators supported on a subregion in $S^{d-1}$.
This means that $\phi_{\rm Rindler}(\rho=0) $ of the AdS-Rindler reconstruction 
is different from $\phi(\rho=0) $.
%As we have seen, 
%which part of $\phi(\rho=0) $ can be represented by $\phi_{\rm Rindler}(\rho=0) $
%is determined by the consideration of the null-geodesics.

Of course, these discussions of the AdS-Rindler reconstruction are
related to our picture in this paper.
In particular, as we have seen, 
the wave packet reconstructed from a small subregion in CFT
is a part of a bulk local operator reconstructed from a whole space $S^{d-1}$,
i.e. the bulk local operator is obtained by assembling such wave packets for small subregions in whole space.
The null-geodesics correspond to such wave packets.
The AdS-Rindler reconstruction uses only the wave packets corresponding to 
the half of $S^{d-1}$.

The AdS-Rindler reconstruction and our bulk reconstruction picture lead 
another version of the subregion duality, in which 
the (low energy) CFT operators supported in region $A$ is dual to
the part of bulk operators in the bulk theory in the AdS-Rindler patch $M_A$.
%with the boundary condition on
%the AdS-Rindler horizon.
We stress again that these are not equivalent to the bulk operators supported on AdS-Rindler patch $M_A$
in the bulk theory in the global AdS.
%without imposing the boundary condition on the AdS-Rindler horizon.
By tracing out the states supported in $\bar{A}$, 
this duality  becomes the correspondence between 
the density matrix $\rho_A$ of the region $A$ in CFT and
the density matrix $\rho_{M_A}$ of the AdS-Rindler patch in the ``bulk theory'' 
which includes the part of the bulk local operators.
%with the boundary condition. 
%Here, we assumed that the boundary condition on the AdS-Rindler horizon
%is consistent with the CFT defined on $A$.
%This duality for the density matrices for general $A$ is often referred to as the subregion duality and
%our picture is consistent with this version of the subregion duality.

\subsection{Disjoint regions }

\begin{figure}
  \centering
  \includegraphics[width=5cm]{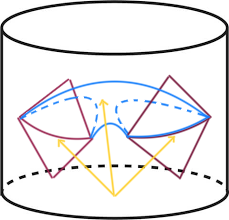}
  \caption{Two intervals (region $A$) and their causal diamonds on the boundary are represented 
by red lines.
Inside the blue curve are the entanglement wedge and inside the dotted blue curves 
are causal wedges.
The two light rays, represented by the yellow lines on the boundary, 
reach the region $A$ at $t=0$. Thus, the operator is supported on the region $A$.
The other yellow line, which represents the bulk picture of them, can reach
the entanglement wedge outside the causal wedges at $t=0$.
}
\label{fig6}
\end{figure}

Now, we will take the subregion $A$ as a space which is (topologically) different from the ball-shaped space.
For the concreteness of the discussion, 
we will take $A$ as a sum of disjoint two intervals $A_1,A_2$  for $d=2$ CFT.
(A ball shaped region is an interval for $d=2$.)
For this case, we can apply the previous discussion on the ball shaped region 
to each $A_i$.
Then, we find that there are
CFT operators 
supported in the region $A=A_1 \cup A_2$ correspond
to bulk operators supported in 
the causal wedge of $A_1$ or $A_2$.
Other than these, there are other kinds of 
CFT operators 
supported in the region $A$. 
Indeed, 
the time evolution of the CFT local operator at $(\Omega_0,t_0)$ to $t=0$ is 
supported in the region $A$
if the intersection of the light cone of $(\Omega_0,t_0)$
and the $t=0$ slice in CFT is contained in $A=A_1 \cup A_2$.
The bulk operators corresponding to such CFT operators
are the wave packets emitting from $(\Omega_0,t_0)$ to the bulk in 
arbitrary directions. 
Then, these bulk operators are supported in the bulk region 
between the two minimal surfaces connecting $A_1$ and $A_2$ 
 (Figure \ref{fig6})
as shown in \cite{local}.

This region coincides with the entanglement wedge \cite{Wa}-\cite{Czech:2012bh}
if the sizes of $A_i$ are sufficiently large compared with the distance between 
$A_1$ and $A_2$.
It had been difficult to understand how to 
reconstruct the bulk operators outside the causal wedge.\footnote{
In \cite{Takayanagi}, this problem was studied although there are some differences between our set up 
and the one in \cite{Takayanagi}. One such difference is that $\Delta \gg 1$ and the bulk local operator is smeared over the small length scale 
$1/\Delta$ in \cite{Takayanagi}. Here, the bulk local operator is smeared over the length scale much smaller than $1/\Delta$, thus the operator considered in \cite{Takayanagi} is regarded as a bulk non-local operator..
}
It is interesting that we can explicitly reconstruct it here.

If the sizes of $A_i$ are sufficiently small, it is believed that
the entanglement wedge becomes the causal wedge and
these bulk operators reconstructed from the CFT operators supported in $A$
are outside the entanglement wedge of $A$.
This phase transition like behavior does not appear in our case
and there seem some contradictions.
However, the Ryu-Takayanagi surface appears in the computation of 
the entanglement entropy of region $A$ using the replica trick in the path-integral, and
there are no concrete arguments that this implies that the relation between the
Ryu-Takayanagi surface and the bulk region in which the 
bulk operators corresponding to $A$ are supported.
Furthermore, it is difficult to determine
in which states 
the bulk modular Hamiltonian, which appears in the relative entropy \cite{Jafferis:2015del},
acts.
Thus, there will be no contradictions between our results and 
the phase transition.\footnote{
The phase transition of the entanglement entropy occurs because 
the exponential enhancement of the contribution of the minimal surfaces
in the large $N$ limit.
However, both in the gravity computation \cite{Lewkowycz:2013nqa} and CFT computation \cite{Hartman}, 
it seems that 
the large $N$ limit is taken first before taking the $n \rightarrow 1$ limit of the 
replica trick and the these limits could not commute.
Indeed, taking the $n \rightarrow 1$ limit first, the exponential suppression
factor for the large $N$ limit
will disappear because the exponent is proportional to $n-1$ also.
As an illustration, $\sum_{i=1}^m e^{ a_i (n-1) N^2  } \rightarrow \sum_{i=1}^m  a_i (n-1) N^2   $
where $a_i$ are constants.
In the $n \rightarrow 1$ limit,  the calculation becomes the infinitely small deformation
of the geometry for which the partition function is proportional to the deformation of the action, not exponential of it.
Thus, the phase transition itself may not be realized if we take 
the $n \rightarrow 1$ limit first, which is the correct way to obtain the 
entanglement entropy in the CFT which is dual to the quantum gravity on AdS.
A similar modification of the Ryu-Takayanagi formula was proposed in \cite{Tsujimura}.
}

\section{Generalization to asymptotic AdS }

In this section, we will generalize 
our picture of the reconstruction of the bulk local state or operators
to the asymptotic AdS spacetime.
Here, we consider, instead of the vacuum, a CFT state $\ket{g_{\mu \nu}}$ which corresponds to the
semi-classical asymptotic AdS spacetime background of the bulk theory.
First, we will consider a part of the geometry in which the causal structure is trivial
and there is no horizon for simplicity.
We expect that the BDHM relation (\ref{BDHMrelation})
is hold for this case also and we assume it.\footnote{
A generalization of \cite{op} to general classical backgrounds
was done in \cite{classical} and showed that the Einstein equation was derived with assumptions made in \cite{op}. It is interesting to determine the spectrum around 
the background and derive the BDHM relation
in this set up.
}
We also assume as usual that in the large $N$ limit
the fluctuations around this background in the bulk theory is 
free theory on this  asymptotic AdS spacetime.
We only consider this perturbative approximation. 

Let us consider a bulk local state at a point $p$ on a time slice ($t=0$) and 
the backward-time evolution of it.
(Here, the time slice is defined by the dilatation operator, for example.)
As in the AdS background, this evolutions is on
the past light-cone of the point $p$ %where the bulk local operator is inserted
because the local operator contains arbitrary high-momentum modes. 
Then, the bulk local state % on the time slice 
is represented as the bulk local operators acting on $\ket{g_{\mu \nu}}$ integrated over a submanifold $M_p^-$ on the boundary of 
the asymptotic AdS spacetime, where $M_p^-$ is
the intersection of this past  light-cone and the boundary of the asymptotic AdS spacetime.
Using the BDHM relation, we can replace the bulk local operators on the boundary 
as the CFT primary operators.
Thus, the bulk local state at $p$ is reconstructed from 
the CFT primary operators acting on $\ket{g_{\mu \nu}}$ integrated over $M_p$ which 
we regard a subregion in the spacetime for CFT.

Instead of the bulk local state $\phi(x) \ket{g_{\mu \nu}} $, 
we can consider bulk local operator $\phi(x)=\phi(x)^+ + \phi(x) ^-$ itself.
Note that $\phi(x) \ket{g_{\mu \nu}} =\phi(x)^+ \ket{g_{\mu \nu}}$ 
determines $\phi(x)^+$ up to an overall phase in the large $N$ limit.
In order to construct  $\phi(x)=\phi(x)^+ + \phi(x) ^-$, 
we need to give this phase.
Note that the time-evolution is symmetric for past and future.
Thus, as for the AdS spacetime, we expect that 
the bulk local operator is reconstructed from 
the CFT primary operators integrated over $M_p= M_p^+ \cup M_p^-$
where 
$M_p^+$ and $M_p^-$ is for the future and past light cones, respectively.

The bulk operator corresponding to
the wave packet moving along a light-like trajectory 
is also realized by a CFT operator, as for the AdS spacetime.
Indeed, the positive energy parts of the CFT operator $\phi_A(z)$ defined by \eqref{phiA}, which was constructed for AdS spacetime,
represents the wave packet moving into the bulk in the direction determined by the tilt $\delta t(\Omega)$
near $z=0$, i.e. on the boundary
because the near the boundary the BDHM relation is assumed to be hold.
Then, this represents the wave packet moving along the light-like trajectory
of the asymptotic AdS geometry.

The causality and duality should be hold for the above wave packet also.
Assuming the null-energy condition for the bulk theory, 
by the Gao-Wald theorem \cite{Gao-Wald}, 
any ``fastest null geodesic'' connecting two points on the boundary must lie entirely within the boundary.
On the other hand, the wave packets should reach the boundary point at the same time 
in the both bulk and CFT pictures because of the duality and the BDHM relation.
Thus,
the bulk wave packet, which is on a light-like trajectory, should correspond to time-like trajectories on the boundary in the CFT picture.
%in order to be consistent with the duality.
This is explained by the group velocity of the wave packet in the CFT picture is always lower than the speed of light
in the vacuum.
In the CFT picture, the asymptotic AdS spacetime is represented as the non-trivial
CFT state $\ket{g_{\mu \nu}}$ and
the speed of the wave packet will become lower 
in this medium.
Of course, in the bulk picture this effects of medium is included in the metric and the velocity of the 
wave packet in the bulk is the speed of light.
Thus, our picture of the bulk local operators for the asymptotic AdS spacetime is consistent with
the Gao-Wald theorem \cite{Gao-Wald}.\footnote{
In \cite{Gao-Wald} some energy conditions were assumed.
In \cite{Ishibashi:2019nby}, they attempted to understand the bulk energy conditions 
by the causality constraint. 
%In our picture, a bulk wave packet is represented by CFT wave packets moving in particular directions. This property has not seem to be incorporated there.
}

\subsection{Black hole}

Let us consider the bulk reconstruction in
the black hole in the asymptotic AdS spacetime.
First, if we consider a (spherical symmetric) heavy star, then 
a bulk local state or a wave packet at the center of the star at $t=0$
will be reconstructed by CFT primary operators 
integrated on a region in $t < -\pi/2$ because of the time delay.
If the star is nearer the black hole, the time delay is larger and can be arbitrary large, which means the whole past region $(t \leq 0)$ of the CFT space $S^{d-1}$ are needed to reconstruct the bulk local operators at $t=0$.
This means that arbitrary large number of (approximately) independent fields are required
because the bulk local operators at different points should be independent each other and different.
Of course, this is impossible because $N$ is large, but finite although 
$N$ is formally infinite in the large $N$ expansion.

For the black hole, the bulk local states or a wave packets near the horizon,
need infinitely many independent fields.
This implies that the CFT can not reconstruct the bulk fields 
even outside the horizon.
This is the same conclusion by the brick wall \cite{tHooft} \cite{brick}  where some hypothetical boundary conditions were imposed.
Here, even though we have not not imposed such boundary conditions,
we conclude the shortage of the CFT states to reconstruct the bulk fields near the horizon, which 
strongly indicates the existence of the brick wall \cite{tHooft}/fuzzball \cite{Mathur1} \cite{Mathur2}/firewall \cite{AMPS} for the (single sided) black hole.
Note that it is important to include the finite $N$ effects, 
which are the (strong) non-perturbative effects of 
the gravitational coupling from the bulk viewpoint, for this conclusion.
Thus, this should be missed by the semi-classical analysis in the bulk.

%
%Why localization in radial direction appears?
%Why causality in radial direction appears?
%Conformal symmetry will not useful because 
%non-conformal theory will be causal and local
%where only the near-boundary region is related 
%to conformal symmetry.
%
%n 1/N expansion (inclusion of the interaction in AdS), 
%locality and causality will hold.
%Which property of CFT guarantees it?
%Anyway, what is a definition of 1/N expansion in CFT?
%
%What state in CFT does correspond to classical geometry in asymptotic AdS? Why the locality and causality appear?
%
%By the lattice like construction diffeo will always
%appear. but, graviton will not usually. Why?
% 
%Black hole formation?
%
%Coherent states of $a^\dagger$ as semi-classical geometry?
%
%SUSY to KK states
%
%4d N=4 from SQM then, large N limit gives QG?
%
%UV-IR

\section*{Acknowledgments}

S.T. would like to thank  Sinya Aoki, Janos Balog, Yuya Kusuki, Takeshi Morita, Lento Nagano,
Sotaro Sugishita, Kotaro Tamaoka, Tadashi Takayanagi and Masataka Watanabe
for useful discussions.
S.T. would like to thank the Yukawa Institute for Theoretical Physics at Kyoto University.
Discussions
during the YITP workshop YITP-W-20-03 on "Strings and Fields 2020" were useful to complete this work.
This work was supported by JSPS KAKENHI Grant Number 17K05414.

\vspace{1cm}

%\noindent
%{\bf Note added}: 

\appendix

\section{Map between bulk and CFT operators }

\label{A}

In this Appendix, we will review the explicit map between the bulk and CFT operators 
in the energy eigen value basis \cite{local}
and show how the bulk wave packet is represented by the CFT primary operators.
We also 
will see that 
only the CFT operators on $t=\pm\pi/2$ are relevant
in the global HKLL reconstruction.

We expand the quantized bulk field $\phi$ with the spherical harmonics $Y_{lm}(\Omega)$,
\begin{align}
\phi(t,\rho,\Omega)=
\sum_{n,l,m} 
\left( 
\hat{a}_{nlm}^\dagger e^{i \omega_{n l} t}  Y_{lm} ( \Omega)
+\hat{a}_{nlm} e^{-i \omega_{n l} t}  Y^*_{lm} ( \Omega)
\right)
\psi_{nl} (\rho) ,
\end{align}
where $\Omega$ represents the coordinates of $S^{d-1}$ and 
the energy eigen value $\omega_{n l} $ is given by
\begin{align}
\omega_{n l} = 2 n+l +\Delta.
\end{align}
The ``wave function'' for the radial direction is given by
\begin{align}
\psi_{nl} (\rho)=\frac{1}{{\cal N}_{nl}} \sin^l (\rho) \cos^\Delta (\rho)\,
P_n^{l+d/2-1, \Delta-d/2} \left( \cos(2 \rho) \right),
\label{psi}
\end{align}
where $P_n^{\alpha, \beta} (x)$ is the Jacobi polynomial defined by
\begin{align}
P_n^{\alpha, \beta} \left( x \right)=
{(-1)^n \over 2^n n!} (1-x)^{-\alpha} (1+x)^{-\beta}
{d^n \over dx^n}  \left(  (1-x)^{\alpha} (1+x)^{\beta}  (1-x^2)^{n}  \right),
\end{align}
and
the normalization constant ${\cal N}_{nl}$ is given by
\begin{align}
{\cal N}_{nl}=(-1)^n \sqrt{
\frac{ \Gamma(n+l+\frac{d}{2})  \Gamma(n+1+\Delta-\frac{d}{2})   }
{  \Gamma(n+l+\Delta) \Gamma(n+1) }
}.
\end{align}
Note that
$(-1)^n {\cal N}_{nl} \rightarrow 1 $.
for $n \rightarrow \infty$ with a fixed $l$.
We will decompose the local operator in the bulk description 
to positive and negative energy modes as
\begin{align}
\phi(t,\rho,\Omega) 
=
\phi^+(t,\rho,\Omega)
+
\phi^-(t,\rho,\Omega),
\end{align}
where $\phi^-(t,\rho,\Omega) =(\phi^+(t,\rho,\Omega))^\dagger$.
%We will concentrate on the one particle states in the large $N$ limit only.

Let us consider the scalar primary field ${\cal O}_\Delta (x)$,  corresponding to $\phi(t,\rho,\Omega) $,
in 
$CFT_d$
on $\mathbf{R} \times S^{d-1}$ where $\mathbf{R}$ is the time direction and 
the radius of $S^{d-1}$ is taken to be one.
For a review of the $CFT_d$, see for example, \cite{Qu, Ry,SD}.
We define the operator 
\begin{align}
\hat{{\cal O}}_\Delta=\lim_{x \rightarrow 0} 
{\cal O}^+_\Delta (x),
\end{align}
where
${\cal O}^+_\Delta (x)$
is the regular parts of ${\cal O}_\Delta (x)$ in $x^\mu \rightarrow 0$ limit\footnote{
This should be done after taking the large $N$ limit.
More precisely,  $\hat{{\cal O}}_\Delta$ is the sum of the operators 
of dimension $\Delta$ up to $1/N$ corrections in ${\cal O}_\Delta (x)$.
}
which can be expanded by the polynomial of $x^\mu$.

The identification of the CFT states
to the states of the Fock space of  
the scalar fields in AdS
is explicitly given by the identification of the creation operators as
\begin{align}
\hat{a}_{n l m}^\dagger =
c_{nl} \, s^{\mu_1 \mu_2 \ldots \mu_l}_{(l,m)} P_{\mu_1} P_{\mu_2} \cdots P_{\mu_l} 
 (P^2)^n \hat{{\cal O}}_\Delta
\label{id1}
\end{align}
where $c_{nl}$ is the normalization constant,
which was given in \cite{op},
the translation operation $P^\mu$ act on an operator such that
$P^\mu \phi =[\hat{P}^\mu, \phi]$
and $s^{\mu_1 \mu_2 \ldots \mu_l}_{(l,m)}$ is
the normalized rank $l$ symmetric traceless constant tensor
such that $Y_{lm}(\Omega)=s^{\mu_1 \mu_2 \ldots \mu_l}_{(l,m)} x_{\mu_1} x_{\mu_2} \cdots x_{\mu_l} /|x|^l$.

Then,
the bulk local operator is expressed as
\begin{align}
{\phi}^+(t=0,\rho,\Omega)
&=\sum_{n,l,m}  
\psi_{nl} (\rho)  Y_{lm} ( \Omega)
\hat{a}_{nlm}^\dagger
\nonumber \\ 
&= \sum_{n,l,m}  
\psi_{nl} (\rho)  Y_{lm} ( \Omega)
c_{nl} s^{\mu_1 \mu_2 \ldots \mu_l}_{(l,m)} P_{\mu_1} P_{\mu_2} \cdots P_{\mu_l}  
(P^2)^n  \hat{{\cal O}}_\Delta,
\label{local}
\end{align}
where only the CFT operators appear 
in the last line.
It is convenient to
write both of the bulk and CFT operators in the creation and annihilation operators,
which are energy eigen operators.
Indeed, from
the BDHM relation \cite{BDHM} in our normalization:
\begin{align}
\lim_{\rho \rightarrow \pi/2} 
{ \phi(\rho,\Omega)
\over \cos^\Delta(\rho)}
&= 
\sqrt{\pi \over 2}
\sqrt{ \Gamma(\Delta) \over \Gamma(\Delta+1-d/2) \Gamma(d/2)}
{\cal O}_\Delta (\Omega),
\end{align}
we find 
\begin{align}
{\cal O}_\Delta (\Omega,t ) =
\sum_{n,l,m}
\psi_{nl}^{CFT}
Y_{lm} ( \Omega) e^{i \omega_{nl} t}
\hat{a}_{nlm}^\dagger +h.c.= 
\sum_{n,l,m}
\psi_{nl}^{CFT}
 \left(
Y_{lm} ( \Omega) e^{i \omega_{nl} t} \hat{a}_{nlm}^\dagger 
+ Y^*_{lm} ( \Omega) e^{-i \omega_{nl} t} \hat{a}_{nlm} \right),
\label{CFTop}
\end{align}
where
\begin{align}
\psi_{nl}^{CFT}
\equiv
\sqrt{{2 \over \pi} {\Gamma(d/2)  \over \Gamma(\Delta) \Gamma(\Delta+1-d/2) } }
\sqrt{ \Gamma(n+\Delta+1-d/2) \Gamma(n+l+\Delta)
 \over  \Gamma(n+1) \Gamma(n+l+d/2) },
\end{align}
which reduces to a constant $\psi_{nl}^{CFT} =
\sqrt{{2 \over \pi}   }$  for $\Delta=d/2$.
For $n \rightarrow \infty$ with a fixed $l$, we have 
$\psi_{nl}^{CFT} \rightarrow 
\sqrt{{2 \over \pi} {\Gamma(d/2)  \over \Gamma(\Delta) \Gamma(\Delta+1-d/2) } }
{n^{\Delta-d/2}}$. % up to a $n$-independent overall constant.

Below we will see the relation between the bulk local operator at the origin and the CFT primary operator
integrated over $S^{d-1}$.
%If the bulk scalar saturates the Breitenlohner-Freedman bound \cite{BF}, the CFT scalar has $\Delta=d/2$.
%For this case, we find that
These two operators are given by
\begin{align}
{\phi}(t=0,\rho=0) &=
\sum_{n,l,m} 
\left( 
\hat{a}_{nlm}^\dagger Y_{lm} ( \Omega)
+\hat{a}_{nlm} Y^*_{lm} ( \Omega)
\right)
\psi_{nl} (0)   \\
%=  \sum_{n} 
%(-1)^n 
%\frac{\Gamma(n+d/2) }{\Gamma(n+1)\Gamma(d/2) {\cal N}_{nl} }
%\left( 
%\hat{a}_{n00}^\dagger 
%+\hat{a}_{n00} 
%\right)
&=
\sum_{n} 
(-1)^n 
\sqrt{
\frac{\Gamma(n+\Delta) \Gamma(n+d/2) }{\Gamma(n+1)\Gamma(n+1+\Delta-d/2) }
}
\left( 
\hat{a}_{n00}^\dagger 
+\hat{a}_{n00} 
\right),
\label{a14}
\end{align}
and 
\begin{align}
\int d \Omega \, {\cal O}_\Delta (\Omega,t ) &=
{ 2 \pi^{\frac{d}{2}} \over  \Gamma(d/2) }
\sum_{n}
\psi_{n0}^{CFT}
 \left(
e^{i (2n+\Delta) t} \hat{a}_{n00}^\dagger 
+  e^{-i (2n+\Delta) t} \hat{a}_{n00} \right),
\end{align}
where we have used $Y_{00}(\Omega)=1$ and
$ \int d \Omega = 2 \pi^{d/2}/ \Gamma(d/2)$.
In particular, for $t=\pm \pi/2$, the latter becomes
\begin{align}
\int d \Omega \, {\cal O}_\Delta (\Omega,t=\pm \pi/2 ) = &
\sum_{n}
 \sqrt{8 \pi^{d-2}   \over \Gamma(d/2)  \Gamma(\Delta) \Gamma(\Delta+1-d/2) } 
\sqrt{ \Gamma(n+\Delta) \Gamma(n+\Delta+1-d/2)
 \over  \Gamma(n+1) \Gamma(n+d/2) } \nonumber \\
& \times (-1)^n  \left(
e^{ \pm i\frac{\pi}{2}  \Delta } \hat{a}_{n00}^\dagger 
+  e^{\mp i  \frac{\pi}{2}  \Delta } \hat{a}_{n00} \right),
\end{align}
Comparing these two expressions,
we find
\begin{align}
\phi^+ (t=0,\rho=0) =
C_\Delta  e^{\mp i  \frac{\pi}{2}  \Delta }
\int d \Omega \left( F_\Delta (\partial_t)
\, {\cal O}^+_\Delta (\Omega,t) \right)|_{t=\pm \frac{\pi}{2}},   
\label{cm}
\end{align}
where
\begin{align}
& C_\Delta  = \sqrt{ \Gamma(d/2)  \Gamma(\Delta) \Gamma(\Delta+1-d/2) 
 \over 8 \pi^{d-2}   },\\
& F_\Delta (x) = { \Gamma((-i x-\Delta+d)/2)
 \over \Gamma((-i x+\Delta-d+2)/2) }.
\label{dof}
\end{align}
We have used $  F_\Delta (\partial_t) e^{i (2n+\Delta) t} =  {\Gamma(n+d/2)
 \over \Gamma(n+\Delta-d/2+1)}  e^{i (2n+\Delta) t}$. On other hand,  we find
$  F_\Delta (\partial_t) e^{-i (2n+\Delta) t} = { \Gamma(-n-\Delta+d/2)
 \over \Gamma(-n-d/2+1) } e^{-i (2n+\Delta) t}$
for the annihilation modes.

From \eqref{cm}, we obtain the following relation between the 
the bulk local state and the CFT state:
\begin{align}
\phi (t=0,\rho=0) \ket{0}=
C_\Delta  e^{\mp i  \frac{\pi}{2}  \Delta }
\int d \Omega \left( F_\Delta (\partial_t)
\, {\cal O}_\Delta (\Omega,t) \right)|_{t=\pm \frac{\pi}{2}}  \ket{0}. 
\label{cms}
\end{align}
This relation can be simplified for $\Delta=d-1$, which implies $F_{\Delta} (x)=1$, as
\begin{align}
\phi (t=0,\rho=0) \ket{0}=
\sqrt{ \Gamma(d/2)^2  \Gamma(d-1)  
 \over 8 \pi^{d-2}   } 
e^{\mp i  \frac{\pi}{2}  (d-1) }
\int d \Omega \,
 {\cal O}_\Delta (\Omega,t=\pm \pi/2)  \ket{0}.
\end{align}
This expression, which implies \eqref{sc} and the picture given in Figure \ref{fig2},
was obtained in \cite{local}.
Similarly, for $\Delta=d-1- m$ where $m$ is a positive integer, the relation can be also explicitly expressed as the integration localized on $t=\pm \pi/2$   because $F_{\Delta} (x)$ is a polynomial:
\begin{align}
 F_{\Delta=d-1-m} (x) =  \prod_{m'=0}^{m-1}  ((-i x-1+m-2m')/2).
\end{align}
In \cite{Aoki}, the corresponding results for the bulk local operators were obtained 
using the HKLL bulk reconstruction formula.
For $\Delta=d-1+ m$ where $m$ is a positive integer, we find the following simple relation:
\begin{align}
 F^{-1}_\Delta (\partial_t) \phi (t=0,\rho=0) \ket{0}=
C_\Delta  e^{\mp i  \frac{\pi}{2}  \Delta }
\int d \Omega 
\,  {\cal O}_\Delta (\Omega,t=\pm \pi/2)  \ket{0},
%\label{cms}
\end{align}
where $ F^{-1}_\Delta (x)= \prod_{m'=0}^{m-1}  ((-i x-1+m-2m')/2)$ is a polynomial of $x$.

We will consider the bulk reconstruction for generic values of $\Delta$ and see that 
the bulk local state is represented as the CFT primary operators only on $t=\pm \pi/2$  as in Figure \ref{fig25} for these cases also, even though $F_\Delta$ may contain the infinitely many derivatives.\footnote{
The infinitely many derivatives do not always  mean  non-locality.
As an example, let us consider a delta-function on $S^1$ in momentum space, $e^{i n \theta}$, and 
the derivative operator $(1/(1-(\partial_\theta)^2))^\eta$ acting on it,
$(1/(1+n^2))^\eta e^{i n \theta}$,
where $\eta > 0$.
With a momentum cut-off $\Lambda$, the inner product of these with different $\theta, \theta'$,
$\sum_{n=-\Lambda}^{\Lambda} e^{-i n \theta'} (1/(1+n^2))^\eta e^{i n \theta}$, is ${\cal O} (\Lambda^{0} )$ for $|\theta-\theta'| \gg 1/\Lambda$.
It is 
${\cal O} (\Lambda^{1-2 \eta} )$ for $|\theta-\theta'| \lessapprox 1/\Lambda$ which is larger than $\Lambda^{1/2}$ if $\eta<1/4$.
Thus, 
for $\eta<1/4$ 
the derivative operator keeps the locality effectively for the cut-off theory.
}
We would like to express the right hand side of \eqref{cms}, $\int d \Omega \left( F_\Delta (\partial_t)
\, {\cal O}_\Delta (\Omega,t) \right)|_{t= -\frac{\pi}{2}}  \ket{0}$,
by 
$\int d \Omega 
\, {\cal O}_\Delta (\Omega,t)   \ket{0}$ where $-\pi/2 \leq t < \pi/2$.
In order to do this, 
first, we introduce a UV energy cut-off $\Lambda$ and consider the (1-particle) spherical symmetric state,  $a^\dagger_{n00} \ket{0}$, only for $n=0,1, \ldots, \Lambda$.
Denoting $ {\cal O}_\Delta^{\Lambda} (\Omega',t')$ as
$ {\cal O}_\Delta (\Omega',t')$ with a UV energy cut-off $\Lambda$,
we also introduce 
$ \ket{ {\partial}^q {\cal O}; t} \equiv \int d \Omega ({\partial_t})^q {\cal O}_\Delta^{\Lambda} (\Omega,t) \ket{0}$ 
for $t=t_s \equiv -\frac{\pi}{2} (1-2 s/(\Lambda+1))$ with $s=0,1, \ldots, \Lambda$.
These form a (not-orthonormal) basis of the cut-off spherical symmetric states
for fixed non-negative integer $m, q$. For $q=0$ we will use $\ket{ {\cal O}; t} = \ket{ {\partial}^0 {\cal O}; t} $.
\footnote{
Below, we can take any $q$. in particular, we can take $q=0$ and consider $\ket{  {\cal O}; t} $ only.}
Denoting the inner products of them as
$
g_{s,s'}=
\bra{ 
 {\partial}^q {\cal O}; t_{s'} } 
 {\cal O}; t_s
\rangle$, 
we can express a state $\ket{v}$ in the cut-off space as
$\ket{v} = \sum_{s} \ket{{\cal O}; t_s} \sum_{s'}  (g^{-1})^{s s'} \langle {\partial}^q {\cal O}; t_{s'}   \ket{v}$.

In order to estimate this, we note that, for $x \rightarrow \infty$,  $F(x) \rightarrow (-ix/2)^{-\Delta+d-1}$ which is a power function of $x$.
We also note that 
$\int d \Omega 
\, {\cal O}_\Delta (\Omega,t%=\pm \pi/2
) \sim \sum_n e^{i (2n +\Delta) t} n^{\Delta-d/2} a^\dagger_{n00}  $, where we have approximated the coefficients of $a^\dagger_{n00} $ as the large $n$ leading terms. 
Using these, we find
\begin{align}
g_{s,s'}=
\bra{ 
 {\partial}^q {\cal O}; t_{s'} } 
 {\cal O}; t_s
\rangle
\sim & \sum_{n=0}^\Lambda n^{2 \Delta-d+q} e^{i 2n (t_s -t_{s'}) },  
%\label{ole}
\end{align}
for large $n$.
Thus,  for $|t_s -t_{s'}| \mbox{ mod } \pi \leq 1/ \Lambda$,
we find  $g_{s,s'}= {\cal O} (\Lambda^{2 \Delta-d+q+1} )$, which is divergent  
in the $\Lambda \rightarrow \infty$ limit for $ \Delta>(d-1-q)/2$, 
and $g_{s,s'}= {\cal O} (\Lambda^{0} )$
for $|t_s -t_{s'}| \mbox{ mod } \pi  \gg 1/ \Lambda$.
This implies that $| {\cal O}; t_s \rangle$ is (almost) orthogonal to $| {\cal O}; t_{s'} \rangle$ for $|t_s-t_{s'}| \mbox{ mod } \pi  \gg 1/\Lambda$.
Similarly, 
we find
\begin{align}
\bra{ 
 {\partial}^q {\cal O}; t_{s'} } 
\int d \Omega \left( F_\Delta (\partial_t)
\, {\cal O}_\Delta^\Lambda (\Omega,t) \right)|_{t= -\frac{\pi}{2}}  \ket{0} %\nonumber \\
\sim  \sum_{n=0}^\Lambda n^{ \Delta+q-1} e^{-i 2n (t'+\pi/2) },  %-\frac{n^2}{2 \Lambda^2} }
\label{ole}
\end{align}
which is $ {\cal O} (\Lambda^{\Delta +q} )$  for $|t_{s'}+\pi/2| \mbox{ mod } \pi \leq 1/ \Lambda$, and it is $ {\cal O} (\Lambda^{0} )$
for $|t_{s'}+\pi/2| \mbox{ mod } \pi  \gg 1/ \Lambda$.
Now, we set $\ket{v} =
c_v \int d \Omega \left( F_\Delta (\partial_t)
\, {\cal O}_\Delta^\Lambda (\Omega,t) \right)|_{t= -\frac{\pi}{2}}  \ket{0}$
where the constant $c_v$ is determined by $||\ket{v}||=1$, then we find $c_v= {\cal O} (\Lambda^{(1-d)/2} )$ from \eqref{a14}.
We also introduce $c_b =||\ket{{\cal O}; t_{s}}||^{-1} = {\cal O} (\Lambda^{(d-2\Delta-1)/2} )$, which vanishes in the large $\Lambda$ limit
for $\Delta > (d-1)/2$. Here we assume $\Delta > (d-1)/2$ for simplicity.
Using these estimations, we obtain
$\ket{v} = \sum_{s} c_s ( c_b \ket{{\cal O}; t_s} )$ where 
$c_s$ is ${\cal O} (1)$ for $\pi/2 - |t_{s}| \leq 1/ \Lambda$.
Here, we have used that $c_b \ket{{\cal O}; t_s}$ is (almost) orthonormal.
On the other hand,
$c_s$ is ${\cal O} (\Lambda^{-\Delta})$ for $\pi/2 - |t_{s}| \gg 1/ \Lambda$.
This means that
 $\int d \Omega \left( F_\Delta (\partial_t)
\, {\cal O}_\Delta (\Omega,t) \right)|_{t= -\frac{\pi}{2}}  \ket{0}$
is localized to $t=-\pi/2$.

Instead of the bulk local state, we can reconstruct the bulk local operator.
In order to do this, we first note that 
\begin{align}
& F_\Delta (-x) = F_\Delta (x)
 { \sin ((i x+\Delta-d+2) \pi/2)
 \over \sin ((i x-\Delta+d) \pi /2) },
\end{align}
where we have used $\Gamma(x) \Gamma(-x)=\pi/ \sin (\pi x)$.
This implies  
\begin{align}
  F_\Delta (-\partial_t) e^{i (2n+\Delta) t} = 
 { \sin (d \pi/2)
 \over \sin ((d -2 \Delta)\pi /2) }
 F_\Delta (\partial_t) e^{i (2n+\Delta) t},
\end{align}
and
$  F_\Delta (\partial_t) e^{-i (2n+\Delta) t} = 
 { \sin (d \pi/2)
 \over \sin ((d -2 \Delta)\pi /2) }
 F_\Delta (-\partial_t) e^{-i (2n+\Delta) t} 
$.
We define 
\begin{align}
 F^+_\Delta (x) =
 { \sin ((d -2 \Delta)\pi /2)
 \over \sin (d \pi/2) + \sin ((d -2 \Delta)\pi /2)}
 \left( F_\Delta (x)+ F_\Delta (-x) \right),
%= { \sin ((d -2 \Delta)\pi /2)
% \over 2 \sin ((d -2 \Delta)\pi /2) \cos (\Delta \pi/2)}
% \left( F_\Delta (x)+ F_\Delta (-x) \right).
\end{align}
which satisfies 
$  F^+_\Delta (\partial_t) e^{i (2n+\Delta) t} = 
F_\Delta (\partial_t) e^{i (2n+\Delta) t}
$
and
$  F^+_\Delta (-\partial_t) e^{-i (2n+\Delta) t} = 
F_\Delta (-\partial_t) e^{-i (2n+\Delta) t}
$.
We also define 
\begin{align}
 F^-_\Delta (x) =
 { \sin ((d -2 \Delta)\pi /2)
 \over -\sin (d \pi/2) + \sin ((d -2 \Delta)\pi /2)}
 \left( F_\Delta (x)- F_\Delta (-x) \right),
%= { \sin ((d -2 \Delta)\pi /2)
% \over 2 \sin ((d -2 \Delta)\pi /2) \cos (\Delta \pi/2)}
% \left( F_\Delta (x)+ F_\Delta (-x) \right).
\end{align}
which satisfies 
$  F^-_\Delta (\partial_t) e^{i (2n+\Delta) t} = 
F_\Delta (\partial_t) e^{i (2n+\Delta) t}
$
and
$  F^-_\Delta (-\partial_t) e^{-i (2n+\Delta) t} = 
-F_\Delta (-\partial_t) e^{-i (2n+\Delta) t}
$.
Then, using the following relations,
\begin{align}
\phi^+ (t=0,\rho=0) =
C_\Delta  e^{\mp i  \frac{\pi}{2}  \Delta }
\int d \Omega \left( F_\Delta (\partial_t)
\, {\cal O}^+_\Delta (\Omega,t) \right)|_{t=\pm \frac{\pi}{2}},   
\label{cma}
\end{align}
and
\begin{align}
\phi^- (t=0,\rho=0) =
C_\Delta  e^{\pm i  \frac{\pi}{2}  \Delta }
\int d \Omega \left( F_\Delta (-\partial_t)
\, {\cal O}^-_\Delta (\Omega,t) \right)|_{t=\pm \frac{\pi}{2}},   
\label{cmb}
\end{align}
we find the reconstruction formula of the bulk local operator:
\begin{align}
\phi (t=0,\rho=0) =
\frac{C_\Delta}{  2 \cos ( \frac{\pi}{2}  \Delta )  }
\left(
\int d \Omega \left( F^+_\Delta (\partial_t)
\, {\cal O}_\Delta (\Omega,t) \right)|_{t=\frac{\pi}{2}}
+\int d \Omega \left( F^+_\Delta (\partial_t)
\, {\cal O}_\Delta (\Omega,t) \right)|_{t=-\frac{\pi}{2}}
\right),
\label{cmp}
\end{align}
and
\begin{align}
\phi (t=0,\rho=0) =
\frac{C_\Delta}{  2 \sin ( \frac{\pi}{2}  \Delta )  }
\left(
\int d \Omega \left( F^-_\Delta (\partial_t)
\, {\cal O}_\Delta (\Omega,t) \right)|_{t=\frac{\pi}{2}}
- \int d \Omega \left( F^-_\Delta (\partial_t)
\, {\cal O}_\Delta (\Omega,t) \right)|_{t=-\frac{\pi}{2}}
\right).
\label{cmm}
\end{align}

There are different expressions for the bulk local operator because 
${\cal O}_\Delta (\Omega,t=\pi/2)$  and
${\cal O}_\Delta (\Omega,t=-\pi/2)$ are not independent in the large $N$ limit.
In fact, there are infinitely many expressions which include the 
following one, which only use the CFT primary operators on $t=\pi/2$:
\begin{align}
\phi (t=0,\rho=0) =
C_\Delta
\int d \Omega \left(
\left(
\alpha F_\Delta (\partial_t)
+\bar{\alpha}  F_\Delta (-\partial_t)
\right)
\, {\cal O}_\Delta (\Omega,t) \right)|_{t=\frac{\pi}{2}},
\label{cm3}
\end{align}
and 
the one, which only use the CFT primary operators on $t=-\pi/2$:
\begin{align}
\phi (t=0,\rho=0) =
C_\Delta
\int d \Omega \left(
\left(
\bar{\alpha} F_\Delta (\partial_t)
+\alpha  F_\Delta (-\partial_t)
\right)
\, {\cal O}_\Delta (\Omega,t) \right)|_{t=-\frac{\pi}{2}},
\label{cm4}
\end{align}
where
\begin{align}
\alpha=\frac{\cos (\Delta \pi/2)}{1+ { \sin (d \pi/2)
 \over \sin ((d -2 \Delta)\pi /2) }}
- i \frac{\sin (\Delta \pi/2)}{1- { \sin (d \pi/2)
 \over \sin ((d -2 \Delta)\pi /2) }}, \,\,\,
\bar{\alpha}=\frac{\cos (\Delta \pi/2)}{1+ { \sin (d \pi/2)
 \over \sin ((d -2 \Delta)\pi /2) }}
+ i \frac{\sin (\Delta \pi/2)}{1- { \sin (d \pi/2)
 \over \sin ((d -2 \Delta)\pi /2) }}.
\label{ab}
\end{align}

\subsection{Bulk wave packets  }
\label{A1}

Let us consider the bulk local operator at $\rho=\rho_0$, $\Omega=\Omega_0$ and $t=0$, which is 
${\phi}^+(t=0,\rho_0,\Omega_0)
=\sum_{n,l,m}  
\psi_{nl} (\rho_0)  Y_{lm} ( \Omega_0)
\hat{a}_{nlm}^\dagger$.
If we smear this bulk local operator 
only for the angular direction $\Omega$
for a very small short distance scale $1/M_c$ where $M_c \gg 1$,
we have bulk wave packets moving in the radial direction $\rho$ as
\begin{align}
\phi^{+}_{\rm smear} (\rho_0,\Omega_0)
=\sum_{n,l,m} ^{l < M_c}
\psi_{nl} (\rho_0)  Y_{lm} ( \Omega_0)
\hat{a}_{nlm}^\dagger% +h.c.
,
\end{align}
where 
the smearing was represented by the restriction of the angular momentum $l$ 
because the precise form of the smearing is not important.
The dominant contributions in the summation over $n,l$ 
are those for $M_c \ll n $, which implies $l \ll n$.
The asymptotic behavior of $\psi_{nl} (\rho)$ for large $n$ (with $l$ fixed) is
computed, using the asymptotic behavior of Jacobi polynomial \cite{asymptotic}, as 
\begin{align}
\psi_{nl} (\rho) 
%\rightarrow_{n \rightarrow \infty} 
=
\frac{1}{\sqrt{\pi n}} (\tan z)^{\frac{d-1}{2}} 
\cos \left( (2n+l+\Delta) z- \frac{\pi}{2} (\Delta-\frac{d}{2}+\frac12 )\right) +{\cal O}(n^{-3/2}),
\label{cos1}
\end{align}
where 
\begin{align}
z=\pi/2-\rho,
\end{align}
and the boundary is at $z=0$.
This includes both the ingoing and outgoing waves, which correspond
to the two exponentials in
 $\cos \left( (2n+l+\Delta) z- \frac{\pi}{2} (\Delta-\frac{d}{2}+\frac12 )\right)
=\frac12 e^{ i ( (2n+l+\Delta) z- \frac{\pi}{2} (\Delta-\frac{d}{2}+\frac12 ))}+ 
\frac12 e^{- i ( (2n+l+\Delta) z- \frac{\pi}{2} (\Delta-\frac{d}{2}+\frac12 ))}$.

We will compare this with the same smearing of the CFT primary operator  
${\cal O}_\Delta (\Omega_0,t ) =
\sum_{n,l,m}
\psi_{nl}^{CFT}
Y_{lm} ( \Omega_0) e^{i \omega_{nl} t}
\hat{a}_{nlm}^\dagger +h.c.$ 
for the angular direction $\Omega_0$, i.e.
\begin{align}
{\cal O}^{\rm smear}_\Delta (\Omega_0 ,t ) \equiv
\sum_{n,l,m} ^{l < M_c}
\psi_{nl}^{CFT}
Y_{lm} ( \Omega_0) e^{i \omega_{nl} t}
\hat{a}_{nlm}^\dagger +h.c
\sim
\sum_{n,l,m} ^{l < M_c}
n^{\Delta-d/2}
Y_{lm} ( \Omega_0) e^{i (2n+l+\Delta) t}
\hat{a}_{nlm}^\dagger +h.c.
\label{wp1}
\end{align}
For $0<t<\pi/2$, we find that
this smeared CFT primary operator with the time-derivatives, 
$(\frac{1}{2 i} \frac{\partial }{\partial t})^{-\Delta+d/2-1/2} {\cal O}^{+ \, {\rm smear}}_\Delta (\Omega_0,t )$, is
almost equivalent to the ingoing part of 
a bulk wave packet at $z=t$ and $\Omega=\Omega_0$, i.e. 
%$(\frac1i \frac{\partial }{\partial \rho})^{\Delta-d/2+1/2} {\phi}^{+}_{\rm smear} (\rho ,\Omega_0)|_{\rho=\pi/2-t}$,
$ {\phi}^{+}_{\rm smear} (\rho ,\Omega_0)|_{\rho=\pi/2-t}$,
up to the overall normalizations.\footnote{
\label{f29}
Here, the derivatives $(\frac{1}{2 i} \frac{\partial }{\partial t})^{-\Delta+d/2-1/2}$
is the asymptotic form of a function like the $F_\Delta(x)$ in \eqref{dof},
which includes the contributions we neglected in the large $n$ limit.
In this paper, we do not determine this function, however,
the asymptotic form in the large $n$ limit is enough for reconstruction of the bulk operator corresponding to the local wave packet, as we have seen for the bulk local operator.}
%Here, 
%$(\frac1i  \frac{\partial }{\partial \rho})^{\Delta-d/2+1/2} f(\rho)$, where $f(\rho)$ is %a function of $\rho$, means just the derivatives if $\Delta-d/2+1/2$ is non-negative integer.
%For the other cases, it means that $\int D(\rho'-\rho) f(\rho')$ where
%$D(x)$ is the fourier transformation of $( p)^{\Delta-d/2+1/2} e^{- a^2 p^2}$
%where $a \ll1$. 
%This $D(x)$ is localized on the region $x \ll 1$. % in the approximation neglecting small $a$ limit. 
%Note that here we are considering the asymptotic behaver in the large $n$ (or $p$).}
The differences are negligible if the cut-off $M_c$ (and the cut-off for $n$ which is not explicitly introduced in this paper, but needed for the regularization of the local operator)  is large.
For $0>t>-\pi/2$, it is the outgoing part of 
the bulk wave packet.

Note that the positive energy part of the bulk local operator at the center \eqref{cmp} is represented as
$\int_{S^{d-1}} d\Omega 
(\frac{1}{2 i} \frac{\partial }{\partial t})^{-\Delta+d-1} 
%\left(- \frac{\partial^2 }{\partial t^2} \right)^{\frac{d-\Delta-1}{2}} 
{\cal O}^{\rm +}_\Delta (\Omega ,t )|_{t=\pi/2}$.
On the other hand, for the wave packet, we have 
$ {\phi}^{+}_{\rm smear} (\rho ,\Omega_0)|_{\rho=\pi/2-t} \sim  
(\frac{1}{2 i} \frac{\partial }{\partial t})^{-\Delta+d/2-1/2}
%\left(- \frac{\partial^2 }{\partial t^2} \right)^{\frac{d-\Delta-1}{2}} 
{\cal O}^{\rm smear +}_\Delta (\Omega ,t )$ for $0<t<\pi/2$.
The additional kinematical factor $(\frac{1}{2i} \frac{\partial }{\partial t})^{(1-d)/2}$ is needed because 
the bulk local operator is spherical and the wave packets have definite directions.

\subsection{Bulk wave packet not at the center}
\label{A3}

We will study \eqref{phiA} more explicitly for $d=2$ case.
For this case, the angular direction $\Omega$ is parametrized by $\varphi$
where $-\pi \leq \varphi <\pi$.
Let us take $f_A(\Omega) \sim e^{-\varphi^2/(2  L_{\rm smearing}^2)}$
where $l_{\rm plank} \ll L_{\rm smearing} \ll 1$.
For this small region $A$ around $\varphi=0$, we can approximate $\delta t (\Omega) \sim a \varphi$,
where $a$ is a constant satisfying $|a| <1$ and we will assume $a>0$ for simplicity.
We expand ${\cal O}_\Delta (\Omega)$ in the energy eigen basis
and the spherical harmonics $e^{i \varphi m}$ for $S^1$ where $m \in \mathbf Z$.
Then, the $\Omega$ integration in (\ref{phiA}) for the creation operator is evaluated as
\begin{align} 
\int d \varphi \, e^{-\varphi^2/ (2 L_{\rm smearing}^2) + i a \varphi (2 n +|m|+\Delta) + i \varphi m} \sim e^{-(a (2 n +|m|+\Delta) +m)^2 (L_{\rm smearing} )^2 / 2},
\end{align}
where $n$ is an analogue of the momentum for the radial direction. %, see Appendix \ref{A}.
This is exponentially suppressed 
for large $n$ or $m$ except the following two conditions are satisfied:
$m<0$ and
\begin{align} 
2 a n - (1-a) |m| \ll (L_{\rm smearing} )^{-1}.
\end{align}
For $a=0$ which corresponds to the bulk local operator at the center $\rho_1=0$,
these conditions imply that $m$ is small compared with $(L_{\rm smearing} )^{-1}$, thus the angular momentum is small. 
For $ a  \rightarrow 1$ which corresponds to consider the bulk local operator close to the boundary, 
these conditions imply that $n$ is small, thus the wave packet is along the boundary.
Between the two special cases, 
only the modes which satisfies $2 n/m=-(1-a)/a$ approximately are dominant and
we expect this represents the wave packet along the corresponding light like trajectory
using the asymptotic behaver of the Jacobi polynomials \cite{asymptotic2}.

\subsection{HKLL bulk reconstruction for $d-\Delta <1$ }
\label{A2}

In this subsection, we will consider 
the HKLL reconstruction \eqref{HKLL eq},
\begin{align} 
\phi(\rho=0,t=0)
& \sim \int_{-\frac{\pi}{2}}^{  \frac{\pi}{2} } d t' 
\int_{S^{d-1}}  d \Omega' \, {1 \over (\cos t')^{d-\Delta} } {\cal O}_{\Delta} (\Omega', t') \CR
& \sim 
\sum_{n}
\psi_{n0}^{CFT} \hat{a}_{n00}^\dagger
\int_{-\frac{\pi}{2}}^{  \frac{\pi}{2} } d t' 
\, {1 \over (\cos t')^{d-\Delta} }
 e^{ i (2 n +\Delta) t'}
 +h.c.
,
\label{HKLLa}
\end{align}
for $d = \mbox{odd}$, but not assuming $d-\Delta  \geq 1$,
and will see that only the CFT operators on an arbitrary small region containing $t=\pm\pi/2$ are relevant.
This localization has been seen in Appendix \ref{A} already,
and here, we will confirm it from the different, but equivalent expression \eqref{HKLLa}.

In \eqref{HKLLa}, the $t'$-integration can be written as a Fourier transformation:
\begin{align} 
\int_{-\frac{\pi}{2}}^{  \frac{\pi}{2} } d t' 
\, {1 \over (\cos t')^{d-\Delta} }
 e^{ i (2 n +\Delta) t'}
& = 
\int_{0}^{  2 {\pi} } d x
\, {1 \over 2 |\sin (x/2)|^{d-\Delta} }
 e^{ i (n +\Delta/2) (x-\pi)} \\
&=\int_{0}^{  2 {\pi} } d x
e^{ i  n  (x-\pi)}
\left(
\, {1 \over 2 |\sin (x/2)|^{d-\Delta} }
 e^{ i \frac{\Delta}{2}  (x-\pi)}
\right)
,
\end{align}
where $t'=(x-\pi)/2$.
If we regard the integrand as a periodic function, it is singular at $x=0$.
Note that it is regular if 
$(\Delta-d)/2$ is a non-negative integer 
and $e^{i \pi \Delta}=1$, but these two conditions are not consistent with 
$d = \mbox{odd}$.
Thus, for large $n$ the contribution from the region near $x=0$ is dominant because 
the contributions from
the smooth function are exponentially suppressed for large $n$.
Because the integrand behaves near $x=0$ like $(\Theta(x) + const.) |x|^{\Delta-d}$ for  $e^{i \pi \Delta} \neq 1$
or $|x|^{\Delta-d}$ for  $e^{i \pi \Delta} = 1$, 
the contribution from the region near $x=0$ for large $n$ 
is 
%${\cal O}(1/n)$ for  $e^{i \pi \Delta} \neq 1$ or  
${\cal O}(1/n^{\Delta-d+1})$.
% for  $e^{i \pi \Delta} = 1$.
Combining these with $\psi_{nl}^{CFT} \rightarrow 
{n^{\Delta-d/2}}$,
we find 
%$\phi(\rho=0,t=0)= \sum_{n} {\cal O}(n^{\Delta-d/2-1})  \, \hat{a}_{n00}^\dagger +h.c.$ or 
$\phi(\rho=0,t=0)= \sum_{n} {\cal O}(n^{d/2-1}) \,  \hat{a}_{n00}^\dagger +h.c.$.
%Both of the summations, $\sum_{n} {\cal O}(n^{\Delta-d/2-1}) $ and $\sum_{n} {\cal O}(n^{d/2-1})$,
The summation $\sum_{n} {\cal O}(n^{d/2-1})$
diverges, and then the time integral of \eqref{HKLLa} is localized on an arbitrary small region containing $t= \pm \pi/2$.

For $d = \mbox{even}$ the smearing function includes the $\log \cos t'$ factor
and it is singular at $t'=\pm \pi/2$ as a periodic function.
Then, for this case also, we can see that the discussions for $d = \mbox{odd}$ can be applied.

\section{Commutator of CFT}
\label{B1}

In this appendix, we will see that the propagation of a scalar of a non-trivial CFT in the large $N$ limit is 
light-like as for the free theory.
First, let us consider the $\Delta=d/2$ case.
For this case, the v.e.v of the commutator, only which is relevant in the large $N$ limit, is given in \cite{comm} as
\begin{align}
\bra{0} [\mathcal{O} (t_{1},\Omega_{1}),  \mathcal{O} (t_{2},\Omega_{2})] \ket{0}
& =
\frac{2\pi^{d/2}}{\Gamma(d/2)} \frac{1}{i \sin t_{12} }
\sum_{l=0}^{\infty}
\cos ( (d/2-1+l) t_{12} )
\sum_{m}
Y_{l m}(\Omega_{1})
Y_{l m}(\Omega_{2}),
\label{cd2}
\end{align}
for $e^{2 i t_{12}} \neq 1$.
Comparing with the free scalar $\mathcal{O}^{\mathrm{free}} (t_{1},\Omega_{1})$ commutator on the cylinder \cite{comm}, 
we find 
\begin{align}
\bra{0} [\mathcal{O} (t_{1},\Omega_{1}),  \mathcal{O} (t_{2},\Omega_{2})] \ket{0}
& =
\frac{d-2}{4} \frac{1}{\sin t_{12} }
[{\partial \over \partial t_1} \mathcal{O}^{\mathrm{free}} (t_{1},\Omega_{1}),  \mathcal{O}^{\mathrm{free}} (t_{2},\Omega_{2})],
\label{cd3}
\end{align}
thus up to the $\frac{1}{\sin t_{12} }$ factor, it is same as one for the free scalar. 
This implies that the propagation of a scalar of a non-trivial CFT is 
light-like.\footnote{
The free scalar commutator itself is not non-zero for a time like separation, however,
it is singular on the light cone, which dominates the commutator if we regularize the local operators by a smearing.}

Note that the $\Delta=d/2$ case is only special because of the computational simplicity here,
thus this light-like property is expected to be hold for other value of the conformal dimension $\Delta$.
Indeed, for the primary operator \eqref{CFTop} in our approximation,
we have $\mathcal{O} (t,\Omega)  \ket{0}= e^{-i \pi \Delta} \mathcal{O} (t+\pi,\bar{\Omega})  \ket{0}$.
where $\bar{\Omega}$ is the antipodal point of $\Omega$ in $S^{d-1}$,
This means that the propagation is light-like.


\begin{thebibliography}{999}
\parskip=-2pt


%\cite{tHooft:1993dmi}
\bibitem{holo}
G.~'t Hooft,
``Dimensional reduction in quantum gravity,''
Conf. Proc. C \textbf{930308} (1993), 284-296
[arXiv:gr-qc/9310026 [gr-qc]].


%\cite{Susskind:1994vu}
\bibitem{Susskind}
L.~Susskind,
``The World as a hologram,''
J. Math. Phys. \textbf{36} (1995), 6377-6396
doi:10.1063/1.531249
[arXiv:hep-th/9409089 [hep-th]].


%\cite{Maldacena:1997re}
\bibitem{Maldacena}
  J.~M.~Maldacena,
  ``The Large N limit of superconformal field theories and supergravity,''
  Int.\ J.\ Theor.\ Phys.\  {\bf 38} (1999) 1113
   [Adv.\ Theor.\ Math.\ Phys.\  {\bf 2} (1998) 231]
  doi:10.1023/A:1026654312961
  [hep-th/9711200].
  %%CITATION = doi:10.1023/A:1026654312961;%%




%\cite{Banks:1998dd}
\bibitem{BDHM}
  T.~Banks, M.~R.~Douglas, G.~T.~Horowitz and E.~J.~Martinec,
  ``AdS dynamics from conformal field theory,''
  hep-th/9808016.
  %%CITATION = HEP-TH/9808016;%%






%\cite{Balasubramanian:1998sn}
\bibitem{BKL}
  V.~Balasubramanian, P.~Kraus and A.~E.~Lawrence,
  ``Bulk versus boundary dynamics in anti-de Sitter space-time,''
  Phys.\ Rev.\ D {\bf 59} (1999) 046003
  doi:10.1103/PhysRevD.59.046003
  [hep-th/9805171].
  %%CITATION = doi:10.1103/PhysRevD.59.046003;%%






%\cite{Heemskerk:2009pn}
\bibitem{Pol}
  I.~Heemskerk, J.~Penedones, J.~Polchinski and J.~Sully,
  ``Holography from Conformal Field Theory,''
  JHEP {\bf 0910} (2009) 079
  doi:10.1088/1126-6708/2009/10/079
  [arXiv:0907.0151 [hep-th]].
  %%CITATION = doi:10.1088/1126-6708/2009/10/079;%%



%\cite{Fitzpatrick:2012cg}
\bibitem{FK}
  A.~L.~Fitzpatrick and J.~Kaplan,
  ``AdS Field Theory from Conformal Field Theory,''
  JHEP {\bf 1302} (2013) 054
  doi:10.1007/JHEP02(2013)054
  [arXiv:1208.0337 [hep-th]].
  %%CITATION = doi:10.1007/JHEP02(2013)054;%%


%\cite{Miyaji:2015fia}
\bibitem{Ta}
  M.~Miyaji, T.~Numasawa, N.~Shiba, T.~Takayanagi and K.~Watanabe,
  ``Continuous Multiscale Entanglement Renormalization Ansatz as Holographic Surface-State Correspondence,''
  Phys.\ Rev.\ Lett.\  {\bf 115} (2015) no.17,  171602
  doi:10.1103/PhysRevLett.115.171602
  [arXiv:1506.01353 [hep-th]].
  %%CITATION = doi:10.1103/PhysRevLett.115.171602;%%

%\cite{Nakayama:2015mva}
\bibitem{NO1}
  Y.~Nakayama and H.~Ooguri,
  ``Bulk Locality and Boundary Creating Operators,''
  JHEP {\bf 1510} (2015) 114
  doi:10.1007/JHEP10(2015)114
  [arXiv:1507.04130 [hep-th]].
  %%CITATION = doi:10.1007/JHEP10(2015)114;%%




%\cite{Verlinde:2015qfa}
\bibitem{Ver}
  H.~Verlinde,
  ``Poking Holes in AdS/CFT: Bulk Fields from Boundary States,''
  arXiv:1505.05069 [hep-th].
  %%CITATION = ARXIV:1505.05069;%%



%\cite{Bena:1999jv}
\bibitem{Bena}
  I.~Bena,
  ``On the construction of local fields in the bulk of AdS(5) and other spaces,''
  Phys.\ Rev.\ D {\bf 62} (2000) 066007
  doi:10.1103/PhysRevD.62.066007
  [hep-th/9905186].
  %%CITATION = doi:10.1103/PhysRevD.62.066007;%%





%\cite{Hamilton:2005ju}
\bibitem{HKLL1}
  A.~Hamilton, D.~N.~Kabat, G.~Lifschytz and D.~A.~Lowe,
  ``Local bulk operators in AdS/CFT: A Boundary view of horizons and locality,''
  Phys.\ Rev.\ D {\bf 73} (2006) 086003
  doi:10.1103/PhysRevD.73.086003
  [hep-th/0506118].
  %%CITATION = doi:10.1103/PhysRevD.73.086003;%%


%\cite{Hamilton:2006az}
\bibitem{HKLL}
  A.~Hamilton, D.~N.~Kabat, G.~Lifschytz and D.~A.~Lowe,
  ``Holographic representation of local bulk operators,''
  Phys.\ Rev.\ D {\bf 74} (2006) 066009
  doi:10.1103/PhysRevD.74.066009
  [hep-th/0606141].
  %%CITATION = doi:10.1103/PhysRevD.74.066009;%%




\bibitem{ElShowk:2011ag}
  S.~El-Showk and K.~Papadodimas,
  ``Emergent Spacetime and Holographic CFTs,''
  JHEP {\bf 1210} (2012) 106
  doi:10.1007/JHEP10(2012)106
  [arXiv:1101.4163 [hep-th]].
  %%CITATION = doi:10.1007/JHEP10(2012)106;%%

%\cite{Kabat:2011rz}
\bibitem{Kabat:2011rz}
  D.~Kabat, G.~Lifschytz and D.~A.~Lowe,
  ``Constructing local bulk observables in interacting AdS/CFT,''
  Phys.\ Rev.\ D {\bf 83} (2011) 106009
  doi:10.1103/PhysRevD.83.106009
  [arXiv:1102.2910 [hep-th]].
  %%CITATION = doi:10.1103/PhysRevD.83.106009;%%

%\cite{Kabat:2012hp}
\bibitem{Kabat:2012hp}
  D.~Kabat, G.~Lifschytz, S.~Roy and D.~Sarkar,
  ``Holographic representation of bulk fields with spin in AdS/CFT,''
  Phys.\ Rev.\ D {\bf 86} (2012) 026004
  doi:10.1103/PhysRevD.86.026004, 10.1103/PhysRevD.86.029901
  [arXiv:1204.0126 [hep-th]].
  %%CITATION = doi:10.1103/PhysRevD.86.026004, 10.1103/PhysRevD.86.029901;%%

%\cite{Kabat:2012av}
\bibitem{Kabat:2012av}
  D.~Kabat and G.~Lifschytz,
  ``CFT representation of interacting bulk gauge fields in AdS,''
  Phys.\ Rev.\ D {\bf 87} (2013) no.8,  086004
  doi:10.1103/PhysRevD.87.086004
  [arXiv:1212.3788 [hep-th]].
  %%CITATION = doi:10.1103/PhysRevD.87.086004;%%
  %30 citations counted in INSPIRE as of 13 Sep 2017




%\cite{Fitzpatrick:2014vua}
\bibitem{FKW}
  A.~L.~Fitzpatrick, J.~Kaplan and M.~T.~Walters,
  ``Universality of Long-Distance AdS Physics from the CFT Bootstrap,''
  JHEP {\bf 1408} (2014) 145
  doi:10.1007/JHEP08(2014)145
  [arXiv:1403.6829 [hep-th]].
  %%CITATION = doi:10.1007/JHEP08(2014)145;%%






%\cite{Kabat:2015swa}
\bibitem{Kabat:2015swa}
  D.~Kabat and G.~Lifschytz,
  ``Bulk equations of motion from CFT correlators,''
  JHEP {\bf 1509} (2015) 059
  doi:10.1007/JHEP09(2015)059
  [arXiv:1505.03755 [hep-th]].
  %%CITATION = doi:10.1007/JHEP09(2015)059;%%





%\cite{Kabat:2016zzr}
\bibitem{Kabat:2016zzr}
  D.~Kabat and G.~Lifschytz,
  ``Locality, bulk equations of motion and the conformal bootstrap,''
  JHEP {\bf 1610} (2016) 091
  doi:10.1007/JHEP10(2016)091
  [arXiv:1603.06800 [hep-th]].
  %%CITATION = doi:10.1007/JHEP10(2016)091;%%


%\cite{Goto:2016wme}
\bibitem{Goto}
  K.~Goto, M.~Miyaji and T.~Takayanagi,
  ``Causal Evolutions of Bulk Local Excitations from CFT,''
  JHEP {\bf 1609} (2016) 130
  doi:10.1007/JHEP09(2016)130
  [arXiv:1605.02835 [hep-th]].
  %%CITATION = doi:10.1007/JHEP09(2016)130;%%

\bibitem{Kim:2016ipt}
  J.~W.~Kim,
  ``Explicit reconstruction of the entanglement wedge,''
  JHEP {\bf 1701} (2017) 131
  doi:10.1007/JHEP01(2017)131
  [arXiv:1607.03605 [hep-th]].
  %%CITATION = doi:10.1007/JHEP01(2017)131;%%

\bibitem{Goto2}
  K.~Goto and T.~Takayanagi,
  ``CFT descriptions of bulk local states in the AdS black holes,''
  JHEP {\bf 1710} (2017) 153
  doi:10.1007/JHEP10(2017)153
  [arXiv:1704.00053 [hep-th]].
  %%CITATION = doi:10.1007/JHEP10(2017)153;%%


%\cite{Dobrev:1998md}
\bibitem{Dobrev:1998md}
V.~K.~Dobrev,
``Intertwining operator realization of the AdS / CFT correspondence,''
Nucl. Phys. B \textbf{553} (1999), 559-582
doi:10.1016/S0550-3213(99)00284-9
[arXiv:hep-th/9812194 [hep-th]].



\bibitem{Rehren}
  M.~Duetsch and K.~H.~Rehren,
  ``Generalized free fields and the AdS - CFT correspondence,''
  Annales Henri Poincare {\bf 4} (2003) 613
  doi:10.1007/s00023-003-0141-9
  [math-ph/0209035].
  %%CITATION = doi:10.1007/s00023-003-0141-9;%%




\bibitem{op}
  S.~Terashima,
  ``AdS/CFT Correspondence in Operator Formalism,''
  JHEP {\bf 1802} (2018) 019
  doi:10.1007/JHEP02(2018)019
  [arXiv:1710.07298 [hep-th]].
  %%CITATION = doi:10.1007/JHEP02(2018)019;%%






%\cite{Gubser:1998bc}
\bibitem{GKP}
  S.~S.~Gubser, I.~R.~Klebanov and A.~M.~Polyakov,
  ``Gauge theory correlators from noncritical string theory,''
  Phys.\ Lett.\ B {\bf 428} (1998) 105
  doi:10.1016/S0370-2693(98)00377-3
  [hep-th/9802109].
  %%CITATION = doi:10.1016/S0370-2693(98)00377-3;%%
  %7361 citations counted in INSPIRE as of 13 Sep 2017

%\cite{Witten:1998qj}
\bibitem{W}
  E.~Witten,
  ``Anti-de Sitter space and holography,''
  Adv.\ Theor.\ Math.\ Phys.\  {\bf 2} (1998) 253
  [hep-th/9802150].
  %%CITATION = HEP-TH/9802150;%%





\bibitem{local}
S.~Terashima,
``Bulk Locality in AdS/CFT Correspondence,''
[arXiv:2005.05962 [hep-th]].


%\cite{Bousso:2012mh}
\bibitem{Bousso:2012mh}
R.~Bousso, B.~Freivogel, S.~Leichenauer, V.~Rosenhaus and C.~Zukowski,
``Null Geodesics, Local CFT Operators and AdS/CFT for Subregions,''
Phys. Rev. D \textbf{88} (2013), 064057
doi:10.1103/PhysRevD.88.064057
[arXiv:1209.4641 [hep-th]].



\bibitem{comm}
L.~Nagano and S.~Terashima,
``A Note on Commutation Relation in Conformal Field Theory,''
[arXiv:2101.04090 [hep-th]].



\bibitem{Gao-Wald}
S.~Gao and R.~M.~Wald,
``Theorems on gravitational time delay and related issues,''
Class. Quant. Grav. \textbf{17} (2000), 4999-5008
doi:10.1088/0264-9381/17/24/305
[arXiv:gr-qc/0007021 [gr-qc]].




\bibitem{Berenstein}
D.~Berenstein and D.~Grabovsky,
``The Tortoise and the Hare: A Causality Puzzle in AdS/CFT,''
[arXiv:2011.08934 [hep-th]].





\bibitem{RT}
S.~Ryu and T.~Takayanagi,
``Holographic derivation of entanglement entropy from AdS/CFT,''
Phys. Rev. Lett. \textbf{96} (2006), 181602
doi:10.1103/PhysRevLett.96.181602
[arXiv:hep-th/0603001 [hep-th]]; 
``Aspects of Holographic Entanglement Entropy,''
JHEP \textbf{08} (2006), 045
doi:10.1088/1126-6708/2006/08/045
[arXiv:hep-th/0605073 [hep-th]].



%\cite{Wall:2012uf}
\bibitem{Wa}
A.~C.~Wall,
``Maximin Surfaces, and the Strong Subadditivity of the Covariant Holographic Entanglement Entropy,''
Class. Quant. Grav. \textbf{31} (2014) no.22, 225007
doi:10.1088/0264-9381/31/22/225007
[arXiv:1211.3494 [hep-th]].


%\cite{Headrick:2014cta}
\bibitem{Headrick:2014cta}
M.~Headrick, V.~E.~Hubeny, A.~Lawrence and M.~Rangamani,
``Causality and holographic entanglement entropy,''
JHEP \textbf{12} (2014), 162
doi:10.1007/JHEP12(2014)162
[arXiv:1408.6300 [hep-th]].


%\cite{Czech:2012bh}
\bibitem{Czech:2012bh}
B.~Czech, J.~L.~Karczmarek, F.~Nogueira and M.~Van Raamsdonk,
``The Gravity Dual of a Density Matrix,''
Class. Quant. Grav. \textbf{29} (2012), 155009
doi:10.1088/0264-9381/29/15/155009
[arXiv:1204.1330 [hep-th]].

%\cite{Jafferis:2015del}
\bibitem{Jafferis:2015del}
D.~L.~Jafferis, A.~Lewkowycz, J.~Maldacena and S.~J.~Suh,
``Relative entropy equals bulk relative entropy,''
JHEP \textbf{06} (2016), 004
doi:10.1007/JHEP06(2016)004
[arXiv:1512.06431 [hep-th]].


%\cite{Bousso:2012sj}
\bibitem{Bo}
R.~Bousso, S.~Leichenauer and V.~Rosenhaus,
``Light-sheets and AdS/CFT,''
Phys. Rev. D \textbf{86} (2012), 046009
doi:10.1103/PhysRevD.86.046009
[arXiv:1203.6619 [hep-th]].




%\cite{Polchinski:1999yd}
\bibitem{Polchinski:1999yd}
J.~Polchinski, L.~Susskind and N.~Toumbas,
``Negative energy, superluminosity and holography,''
Phys. Rev. D \textbf{60} (1999), 084006
doi:10.1103/PhysRevD.60.084006
[arXiv:hep-th/9903228 [hep-th]].




\bibitem{Dong:2016eik}
X.~Dong, D.~Harlow and A.~C.~Wall,
``Reconstruction of Bulk Operators within the Entanglement Wedge in Gauge-Gravity Duality,''
Phys. Rev. Lett. \textbf{117} (2016) no.2, 021601
doi:10.1103/PhysRevLett.117.021601
[arXiv:1601.05416 [hep-th]].





\bibitem{Morrison}
I.~A.~Morrison,
``Boundary-to-bulk maps for AdS causal wedges and the Reeh-Schlieder property in holography,''
JHEP \textbf{05} (2014), 053
doi:10.1007/JHEP05(2014)053
[arXiv:1403.3426 [hep-th]].




\bibitem{ADH}
A.~Almheiri, X.~Dong and D.~Harlow,
``Bulk Locality and Quantum Error Correction in AdS/CFT,''
JHEP \textbf{04} (2015), 163
doi:10.1007/JHEP04(2015)163
[arXiv:1411.7041 [hep-th]].







%\cite{KeskiVakkuri:1998nw}
%\bibitem{KV}
%E.~Keski-Vakkuri,
%``Bulk and boundary dynamics in BTZ black holes,''
%Phys. Rev. D \textbf{59} (1999), 104001
%doi:10.1103/PhysRevD.59.104001
%[arXiv:hep-th/9808037 [hep-th]].

%\cite{Sugishita:2022ldv}
\bibitem{Sugishita}
S.~Sugishita and S.~Terashima,
``Rindler Bulk Reconstruction and Subregion Duality in AdS/CFT,''
[arXiv:2207.06455 [hep-th]].


\bibitem{Takayanagi}
Y.~Suzuki, T.~Takayanagi and K.~Umemoto,
``Entanglement Wedges from the Information Metric in Conformal Field Theories,''
Phys. Rev. Lett. \textbf{123} (2019) no.22, 221601
doi:10.1103/PhysRevLett.123.221601
[arXiv:1908.09939 [hep-th]];
%\bibitem{Kusuki:2019hcg}
Y.~Kusuki, Y.~Suzuki, T.~Takayanagi and K.~Umemoto,
``Looking at Shadows of Entanglement Wedges,''
[arXiv:1912.08423 [hep-th]].



%\cite{Lewkowycz:2013nqa}
\bibitem{Lewkowycz:2013nqa}
A.~Lewkowycz and J.~Maldacena,
``Generalized gravitational entropy,''
JHEP \textbf{08} (2013), 090
doi:10.1007/JHEP08(2013)090
[arXiv:1304.4926 [hep-th]].

%\cite{Faulkner:2013ana}
\bibitem{Faulkner:2013ana}
T.~Faulkner, A.~Lewkowycz and J.~Maldacena,
``Quantum corrections to holographic entanglement entropy,''
JHEP \textbf{11} (2013), 074
doi:10.1007/JHEP11(2013)074
[arXiv:1307.2892 [hep-th]].


%\cite{Hartman:2013mia}
\bibitem{Hartman}
T.~Hartman,
``Entanglement Entropy at Large Central Charge,''
[arXiv:1303.6955 [hep-th]].




\bibitem{Tsujimura}
J.~Tsujimura,
``A gravity dual of entanglement entropy,''
[arXiv:2011.00407 [hep-th]].





\bibitem{classical}
S.~Terashima,
``Classical Limit of Large N Gauge Theories with Conformal Symmetry,''
JHEP \textbf{02} (2020), 021
doi:10.1007/JHEP02(2020)021
[arXiv:1907.05419 [hep-th]].





%\cite{Ishibashi:2019nby}
\bibitem{Ishibashi:2019nby}
A.~Ishibashi, K.~Maeda and E.~Mefford,
``Achronal averaged null energy condition, weak cosmic censorship, and AdS/CFT duality,''
Phys. Rev. D \textbf{100} (2019) no.6, 066008
doi:10.1103/PhysRevD.100.066008
[arXiv:1903.11806 [hep-th]].
%\cite{Iizuka:2019ezn}
%\bibitem{Iizuka:2019ezn}
N.~Iizuka, A.~Ishibashi and K.~Maeda,
``Conformally invariant averaged null energy condition from AdS/CFT,''
JHEP \textbf{03} (2020), 161
doi:10.1007/JHEP03(2020)161
[arXiv:1911.02654 [hep-th]].
%\cite{Iizuka:2020wuj}
%\bibitem{Iizuka:2020wuj}
N.~Iizuka, A.~Ishibashi and K.~Maeda,
``The averaged null energy conditions in even dimensional curved spacetimes from AdS/CFT duality,''
JHEP \textbf{10} (2020), 106
doi:10.1007/JHEP10(2020)106
[arXiv:2008.07942 [hep-th]].





\bibitem{tHooft}
  G.~'t Hooft,
  ``On the Quantum Structure of a Black Hole,''
  Nucl.\ Phys.\ B {\bf 256} (1985) 727.
  doi:10.1016/0550-3213(85)90418-3
  %%CITATION = doi:10.1016/0550-3213(85)90418-3;%%




%\cite{Iizuka:2013kma}
\bibitem{brick}
  N.~Iizuka and S.~Terashima,
  ``Brick Walls for Black Holes in AdS/CFT,''
  Nucl.\ Phys.\ B {\bf 895} (2015) 1
  doi:10.1016/j.nuclphysb.2015.03.018
  [arXiv:1307.5933 [hep-th]].
  %%CITATION = doi:10.1016/j.nuclphysb.2015.03.018;%%






%\cite{Mathur1}
\bibitem{Mathur1}
S.~D.~Mathur,
``The Information paradox: A Pedagogical introduction,''
Class. Quant. Grav. \textbf{26} (2009), 224001
doi:10.1088/0264-9381/26/22/224001
[arXiv:0909.1038 [hep-th]].


%\cite{Mathur2}
\bibitem{Mathur2}
S.~D.~Mathur,
``The Fuzzball proposal for black holes: An Elementary review,''
Fortsch. Phys. \textbf{53} (2005), 793-827
doi:10.1002/prop.200410203
[arXiv:hep-th/0502050 [hep-th]].




%\cite{AMPS}
\bibitem{AMPS}
A.~Almheiri, D.~Marolf, J.~Polchinski and J.~Sully,
``Black Holes: Complementarity or Firewalls?,''
JHEP \textbf{02} (2013), 062
doi:10.1007/JHEP02(2013)062
[arXiv:1207.3123 [hep-th]].







\bibitem{Qu}
  J.~D.~Qualls,
  ``Lectures on Conformal Field Theory,''
  arXiv:1511.04074 [hep-th].
  %%CITATION = ARXIV:1511.04074;%%

\bibitem{Ry}
  S.~Rychkov,
  ``EPFL Lectures on Conformal Field Theory in $D>= 3$ Dimensions,''
  doi:10.1007/978-3-319-43626-5
  arXiv:1601.05000 [hep-th].
  %%CITATION = doi:10.1007/978-3-319-43626-5;%%

\bibitem{SD}
  D.~Simmons-Duffin,
  ``The Conformal Bootstrap,''
  doi:10.1142/9789813149441-0001
  arXiv:1602.07982 [hep-th].
  %%CITATION = doi:10.1142/9789813149441_0001;%%





%\cite{Breitenlohner:1982bm}
\bibitem{BF}
  P.~Breitenlohner and D.~Z.~Freedman,
  ``Positive Energy in anti-De Sitter Backgrounds and Gauged Extended Supergravity,''
  Phys.\ Lett.\  {\bf 115B} (1982) 197.
  doi:10.1016/0370-2693(82)90643-8
  %%CITATION = doi:10.1016/0370-2693(82)90643-8;%%

\bibitem{asymptotic}
  Bai, Xiao-Xi and Zhao, Yuqiu,
  ``A uniform asymptotic expansion for Jacobi polynomials via uniform treatment of Darboux's method,''
Journal of Approximation Theory  \textbf{148} (2007) 1
 doi:10.1016/j.jat.2007.02.001

%\cite{Aoki:2021ekk}
\bibitem{Aoki}
S.~Aoki and J.~Balog,
``HKLL bulk reconstruction for small $\Delta$,''
[arXiv:2112.04326 [hep-th]].


\bibitem{asymptotic2}
  Chen, Lichen and Ismail, Mourad,
  ``On Asymptotics of Jacobi Polynomials,''
Siam Journal on Mathematical Analysis - SIAM J MATH ANAL.  \textbf{22} (1991) 
 doi:10.1137/0522092.






\end{thebibliography}
\end{document}